\documentclass[sigconf,nonacm,dvipsnames]{acmart}

\usepackage[T1]{fontenc}
\usepackage[utf8]{inputenc}
\usepackage{algorithm}
\usepackage{algorithmicx}
\usepackage[noend]{algpseudocode}
\usepackage{amsmath}
\usepackage{amssymb}
\usepackage[english]{babel}
\usepackage{enumitem}
\usepackage{float}
\usepackage{hyperref}
\usepackage{listings}
\usepackage{lstautogobble}
\usepackage{subcaption}
\usepackage{tikz}
\usepackage{xspace}
\usepackage{pifont}
\usepackage{balance}

\usetikzlibrary{
    arrows,
    calc,
    chains,
    matrix,
    positioning,
    scopes,
}

\AtBeginDocument{

}

\newfloat{lstfloat}{}{lop}
\floatname{lstfloat}{Listing}

\setlist[enumerate,itemize]{nosep}

\newcommand{\cmark}{\ding{51}}
\newcommand{\xmark}{}
\newcommand{\Cpp}{C\texttt{++}\xspace}
\newcommand{\books}{\lstinline{books}\xspace}
\newcommand{\fb}{\lstinline{fb}\xspace}
\newcommand{\osmc}{\lstinline{osmc}\xspace}
\newcommand{\wiki}{\lstinline{wiki}\xspace}

\newcommand{\ca}[1]{\citeauthor{#1}~\cite{#1}}

\algnewcommand\algorithmicinput{\textbf{Input:}}
\algnewcommand\Input{\item[\algorithmicinput]}
\algnewcommand\algorithmicoutput{\textbf{Output:}}
\algnewcommand\Output{\item[\algorithmicoutput]}

\makeatletter
\algrenewcommand\ALG@beginalgorithmic{\fontsize{7}{7}\selectfont}
\makeatother

\begin{document}
\pagestyle{plain}
\title{A Critical Analysis of Recursive Model Indexes}

\author{Marcel Maltry}
\affiliation{%
  \institution{Saarland Informatics Campus}
}
\email{marcel.maltry@bigdata.uni-saarland.de}

\author{Jens Dittrich}
\affiliation{%
    \institution{Saarland Informatics Campus}
}
\email{jens.dittrich@bigdata.uni-saarland.de}

\begin{abstract}
The \emph{recursive model index} (RMI) has recently been introduced as a machine-learned replacement for traditional indexes over sorted data, achieving remarkably fast lookups. Follow-up work focused on explaining RMI's performance and automatically configuring RMIs through enumeration. Unfortunately, configuring RMIs involves setting several hyperparameters, the enumeration of which is often too time-consuming in practice. Therefore, in this work, we conduct the first inventor-independent broad analysis of RMIs with the goal of understanding the impact of each hyperparameter on performance. In particular, we show that in addition to model types and layer size, error bounds and search algorithms must be considered to achieve the best possible performance. Based on our findings, we develop a simple-to-follow guideline for configuring RMIs. We evaluate our guideline by comparing the resulting RMIs with a number of state-of-the-art indexes, both learned and traditional. We show that our simple guideline is sufficient to achieve competitive performance with other learned indexes and RMIs whose configuration was determined using an expensive enumeration procedure. In addition, while carefully reimplementing RMIs, we are able to improve the build time by 2.5x to 6.3x.
\end{abstract}

\maketitle

\section{Introduction}
\label{sec:intro}

Machine learning and artificial intelligence are taking the world by storm. Research areas that were believed to have been researched to completion have been revisited with exciting new results, showing that considerable improvements are still possible \emph{if} we factor in wisdom from the machine learning world. Notable examples include natural language processing and compute vision which were completely revolutionized in the past decade by variants of deep learning. In the database world, we witnessed a surge of similar re-exploration endeavors in the past five years. Notable recent examples of works in that space include cardinality estimation~\cite{hilprecht19deepdb,woltmann19cardinality}, auto-tuning~\cite{pavlo17self,vanaken17automatic}, and indexing~\cite{kraska2018case,galakatos2019fiting,ding2020alex,ferragina2020pgm,kipf2020radixspline}. We believe that indexing is the most surprising result of these three areas because both cardinality estimation and auto-tuning are optimization problems and thus have a natural proximity to machine learning. The connection to indexing becomes evident when we examine a special case of indexing.

\noindent\textbf{Problem Statement.}
Given a sorted, densely packed array $A$ of keys and a query $Q$ asking for a particular key $x_i$ that may or may not exist in that array, return the array index $i$ of that key $x_i$.

In other words, we are looking for a function that assigns to each key its position in the sorted array. Traditionally, this function is implemented by a suitable algorithm like binary search or a data structure like a B-tree. In contrast, \ca{kraska2018case} observe that this function can be learned through regression, effectively making the indexing problem a machine learning task. Based on this observation, \ca{kraska2018case} present the \textit{recursive model index}~(RMI) as first \textit{learned index} with remarkable results in terms of lookup performance. We wanted to understand the performance benefits of RMIs early on and therefore tried to reproduce the results. However, we quickly encountered several issues.

\noindent\textbf{Hyperparameter configuration.}
Configuring RMIs involves setting several hyperparameters. Unfortunately, the exact configurations with which the remarkable results were obtained were not reported and in some cases even described misleadingly. The use of neural networks is mentioned frequently throughout the experimental evaluation of the original paper. However, the low model evaluation times reported in Fig.~4 strongly suggest that none of the best-performing configurations actually uses neural networks. In personal communication with the first author in August 2019, we learned that linear models should be preferred over neural networks in most cases. In our experience, there is still a misconception in the community today that RMIs internally use neural networks. Subsequent studies~\cite{kipf2019sosd,marcus2020benchmarking} involving inventors of the RMI investigated the performance benefits of learned indexes over traditional indexes. However, hyperparameter configurations for the reported results were obtained by a time-consuming enumeration process~\cite{marcus2020cdfshop}. As a result, similar to the original paper~\cite{kraska2018case}, the studies neither show how the choice of hyperparameters affects performance, nor do they give advice for configuring RMIs in practice besides enumeration.

\noindent\textbf{Closed source.}
The source code of the original paper was never made available. A so-called reference implementation~\cite{marcus2020benchmarking} was published in December 2019, two years after the preprint~\cite{kraska2018casepre}. However, that implementation differs from the descriptions in the original paper in terms of model types and error bounds.

\vspace{.1cm}

\noindent\textbf{Goals:}
We pursue the following objectives with this paper.
\begin{itemize}[leftmargin=*]
    \item Conduct the first inventor-independent detailed analysis of RMIs to understand the impact of each hyperparameter on prediction accuracy, lookup time, and build time.
    \item Develop a clear and simple guideline for database architects on how to configure RMIs with good lookup performance.
    \item Provide a clean and easily extensible implementation of RMIs.
\end{itemize}

\noindent\textbf{Contributions:}
We make the following contributions to achieve these goals.

\noindent\textbf{(1)~Learned Tree-Structured Indexes.}
We revisit in detail recursive model indexes~\cite{kraska2018case} and explain how they are trained and what hyperparameters to consider~(\autoref{sec:rmi}). We provide a detailed overview on the design dimensions of learned indexes and the already large body of work in that space~(\autoref{sec:related}).

\noindent\textbf{(2)~Hyperparameter Analysis.}
We present our experimental setup~(\autoref{sec:setup}) and conduct a set of extensive experiments to analyze the impact of each hyperparameter on predictive accuracy and search interval size~(\autoref{sec:error}), lookup performance~(\autoref{sec:lookup}), and build time~(\autoref{sec:build}).

\noindent\textbf{(3)~Configuration Guideline.}
Based on our findings, we develop a simple guideline to configure RMIs in practice~(\autoref{sec:guideline}).

\noindent\textbf{(4)~Comparison with Other Indexes.}
We compare the RMIs resulting from our guideline in terms of lookup time and build time with a number of learned indexes like ALEX~\cite{ding2020alex}, PGM-index~\cite{ferragina2020pgm}, RadixSpline~\cite{kipf2020radixspline}, and the reference implementation of RMIs~\cite{marcus2020cdfshop}, as well as state-of-the-art traditional indexes like B-tree~\cite{bayer72organisation}, ART~\cite{leis2013adaptive}, and Hist-Tree~\cite{crotty2021histtree}~(\autoref{sec:comparison}).

\section{Recursive Model Indexes}
\label{sec:rmi}

In this section, we recap recursive model indexes, how to perform a lookup, how they are trained, and what their hyperparameters are.

\subsection{Core Idea}
\label{subsec:rmi:core}

RMIs are based on the observation that the position of a key in a sorted array can be computed using the \emph{cumulative distribution function} (CDF) of the data. Let $D$ be a dataset consisting of $n=|D|$ keys. Further, let $X$ be a random variable that takes each key's value with equal probability and let $F_X$ be the CDF of $X$. Then, the position $i$ of each key $x_i \in D$ in the sorted array is computed as:
\begin{equation}
    i = F_X(x_i) \cdot n = P(X \leq x_i) \cdot n
    \label{eq:rmi:idx}
\end{equation}
Note that in the context of learned indexes, the term CDF is frequently used synonymously for a mapping from key to position in the sorted array instead of its statistical definition of a mapping from key to the probability that a random variable will take a value less than or equal to that key. In the following, we submit to the former interpretation.

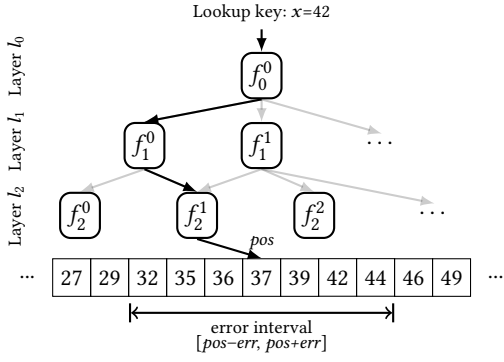
\begin{figure}
    \centering
    \begin{tikzpicture}[node distance=10mm]
    \tikzset{
        model/.style={rectangle,rounded corners,draw,thick},
        arrow/.style={-latex,thick},
        conn/.style={-latex,thick,opacity=0.2},
        array/.style={matrix of math nodes,column sep=-\pgflinewidth,nodes={draw,minimum size=5mm}}
    }

    \node[model]                                (f00) {$f^0_0$};
    \node[model,below left=3mm and 10mm of f00] (f10) {$f^0_1$};
    \node[model,right=of f10]                   (f11) {$f^1_1$};
    \node[right=of f11]                         (f1d) {$\cdots$};
    \node[model,below left=3mm and 3mm of f10]  (f20) {$f^0_2$};
    \node[model,right=of f20]                   (f21) {$f^1_2$};
    \node[model,right=of f21]                   (f22) {$f^2_2$};
    \node[right=of f22]                         (f2d) {$\cdots$};
    \node[left=8mm of f20,rotate=90,anchor=north]  (l2) {\footnotesize Layer $l_2$};
    \node[above=10mm of l2,rotate=90,anchor=north] (l1) {\footnotesize Layer $l_1$};
    \node[above=10mm of l1,rotate=90,anchor=north] (l0) {\footnotesize Layer $l_0$};

    \node[above=3mm of f00]                     (key) {\footnotesize Lookup key: $x$=42};

    \matrix (arr) [array,below=20mm of f00]{
        27 & 29 & 32 & 35 & 36 & 37 & 39 & 42 & 44 & 46 & 49 \\
    };
    \node[left=0mm of arr]  (ldots) {...};
    \node[right=0mm of arr] (rdots) {...};

    \node[below=2mm of arr-1-6]                    (int) {\footnotesize error interval};
    \node[below=4mm of arr-1-6]                    (bounds) {\footnotesize [\textit{pos--err}, \textit{pos+err}]};
    \node[below=1mm of arr-1-2,minimum width=5mm]  (lint) {};
    \node[below=1mm of arr-1-10,minimum width=5mm] (rint) {};
    \node[above=0mm of arr-1-6]                    (pos) {\footnotesize \textit{pos}};
    \draw[|<->|,thick] (lint) -- (rint);

    \draw[arrow] (key) -- (f00);
    \draw[arrow] (f00.south) -- (f10.north);
    \draw[arrow] (f10.south) --(f21.north);
    \draw[arrow] (f21.south) -- (arr-1-6.north);

    \draw[conn]  (f00.south) -- (f10.north);
    \draw[conn]  (f00.south) -- (f11.north);
    \draw[conn]  (f00.south) -- (f1d.north);
    \draw[conn]  (f10.south) -- (f20.north);
    \draw[conn]  (f10.south) -- (f21.north);
    \draw[conn]  (f11.south) -- (f21.north);
    \draw[conn]  (f11.south) -- (f22.north);
    \draw[conn]  (f11.south) -- (f2d.north);
\end{tikzpicture}
    \caption{A three-layer RMI that is evaluated on key~42 yielding estimated position \textit{pos} (prediction). Based on \textit{pos}, the sorted array is searched for the key (error correction).}
    \label{fig:rmi-structure}
\end{figure}

The core idea of an RMI is to \textit{approximate} the CDF of a dataset by means of a hierarchical, multi-layer model. Consider \autoref{fig:rmi-structure} for an example three-layer RMI. Each model in an RMI approximates a segment of the CDF, all models of a layer together approximate the entire CDF. An RMI is a \emph{directed acyclic graph} (DAG), i.e.,~in contrast to a tree, a node (or model) in an RMI may have multiple direct predecessors. We denote the $i$-th layer of a $k$-layered RMI by $l_i$ where $0\leq i \leq k-1$ and refer to the $j$-th model of the $i$-th layer by $f^j_i$. The first layer $l_0$ of an RMI always consists of a single \emph{root model} $f^0_0$. Each subsequent layer may consist of an arbitrary number of models. The number of models of a layer $l_i$ is denoted by $|l_i|$ and called the size of the layer.

\subsection{Index Lookup}
\label{subsec:rmi:lookup}

A lookup is performed in two steps: (1)~\emph{Prediction}: We evaluate the RMI on a given key yielding a position estimate. (2)~\emph{Error correction}: We search the key in the area around the estimated position in the sorted array to compensate for estimation errors. We discuss both steps in more detail below.

\noindent\textbf{Prediction.}
Consider again \autoref{fig:rmi-structure} that shows an index lookup for key 42. We start by evaluating the root model $f^0_0$ on key 42 yielding a position estimate. Based on this estimate, model $f^0_1$ in the next layer $l_1$ is chosen for evaluation. This iterative process is continued until the position estimate \textit{pos} of the last layer is obtained.

\vspace{.1cm}
\noindent\emph{Definition (Prediction)}:
Let $R$ be a $k$-layer RMI trained on dataset $D$ consisting of $n=|D|$ keys. Let us denote the value $p$ restricted to the interval $[a,b]$ by
\begin{equation}
    \big\llbracket p\big\rrbracket_a^b := \text{max}\big(a, \text{min}(p, b)\big).
    \label{eq:clamp}
\end{equation}
The predicted position for key $x$ of layer $l_i$ is recursively defined as
\begin{equation}
    f_i(x) =
    \begin{cases}
        f^0_0(x) & i = 0 \\[5pt]
        f^{\left\lfloor \left\llbracket |l_i| \cdot f_{i-1}(x) / n \right\rrbracket_0^{|l_i|-1} \right\rfloor}_i(x) & 0<i<k \\
    \end{cases}
    \label{eq:fi}
\end{equation}
    Intuitively, to determine the model in layer $l_i$ that is evaluated on key $x$, the estimate $f_{i-1}(x)$ of the previous layer is scaled to the size of the current layer. Note that $f_{i-1}(x)$ might be less than 0 or greater than $n-1$. Thus, the result is restricted to $[0,|l_i|-1]$ to evaluate to a valid model index. The predicted position for key $x$ of RMI $R$ is the output of layer $l_{k-1}$:
\begin{equation}
    R(x) = f_{k-1}(x).
\end{equation}

\noindent\textbf{Error correction.}
Based on the estimate $R(x)$ obtained by evaluating the RMI, the sorted array is searched for the key. In \autoref{fig:rmi-structure}, the position estimate for key 42 points to key 37 in the sorted array. Since 37$<$42, we have to search to the right of 37 to find 42. To facilitate the search, an RMI may store error bounds that limit the size of the interval that has to be searched. The RMI guarantees that if a key is present, then it can be found within the provided error bounds. A simple way of achieving this is to store the maximum absolute error \textit{err} of the RMI. The left and right search bounds, i.e.,~the error interval, is set to $[R(x)-\textit{err},R(x)+\textit{err}]$. If the key exists, it must be within these bounds. We search the interval for key $x$ using an appropriate algorithm like binary search.

\subsection{Training Algorithm}
\label{subsec:rmi:training}

The goal of the training process is to minimize the prediction error. The training algorithm is shown in \autoref{algo:train}. Its core idea is to perform a top-down layer-wise bulk loading. We start by assigning all keys to the root model (line~\ref{init_keys00}). Then, the root model is trained on those keys (line~\ref{train}). Afterwards, the keys are assigned to the next-layer models based on the root model's estimates (lines~\ref{assign-begin}--\ref{assign-end}). We proceed by training the models of the next layer on the keys that were assigned to them. This process is repeated for each layer until the last layer has been trained. Finally, if desired, error bounds can be computed on the trained RMI (after line~\ref{assign-end}).

\begin{lstfloat}[t]
    \caption{RMI Training Algorithm.}
    \label{algo:train}
    \begin{algorithmic}[1]
        \Input Dataset $D$, number of layers $k$, array of layer sizes $l$
        \Output RMI $R$
        \Procedure{BuildRMI}{$D$, $k$, $l$}
            \State $R := \text{Array2D}()$ \Comment{Initialize dynamic array to store models.}
            \State $\textit{keys} := \text{Array2D}()$ \Comment{Initialize dynamic array to store each model's keys.}
            \State $\textit{keys}[0, 0] := D$ \label{init_keys00} \Comment{Assign all keys to the root model.}
            \For{$i \gets 0$ \textbf{to} $k-1$}
                \For{$j \gets 0$ \textbf{to} $l[i]-1$}
                    \State $R[i, j] :=$ \Call{TrainModel}{\textit{keys}[i, j]} \Comment{Train model $j$ of layer $i$.} \label{train}
                    \If{$i < k-1$} \Comment{Check whether current layer is not last layer.}
                        \ForAll{$x$ \textbf{in} $\textit{keys}[i, j]$} \label{assign-begin}
                            \State $p :=$ \Call{GetModelIndex}{$x, R[i, j], l[i+1], |D|$}
                            \State $\textit{keys}[i+1, p].\text{add}(x)$ \Comment{Assign key $x$ to next-layer model $p$.} \label{assign-end}
                        \EndFor
                    \EndIf
                \EndFor
            \EndFor
            \State \Return $R$
        \EndProcedure
        \Statex
        \Function{GetModelIndex}{$x, f, q, n$}
            \State \Return $\left\lfloor \llbracket q \cdot f(x) / n \rrbracket_0^{q-1} \right\rfloor$ \Comment{Compute model index according to Equation~(\ref{eq:fi}).}
        \EndFunction
    \end{algorithmic}
\end{lstfloat}

\subsection{Hyperparameters}
\label{subsec:rmi:hyper}

RMIs offer a high degree of freedom in configuration and tuning. In the following, we briefly describe each hyperparameter. We provide a set of possible configurations for each parameter in \autoref{subsec:setup:hyper} when describing the experimental setup.

\noindent\textbf{Model types.}
Model types are crucial to the predictive quality of RMIs. While simple models, e.g., linear regression, are small and fast to train and evaluate, complex models, e.g., neural networks, might offer higher accuracy, but are slow to train and evaluate.

\noindent\textbf{Layer count.}
The number of layers $k$ determines the depth of an RMI. While a deeper RMI might distribute the keys more evenly over the last-layer models, deeper RMIs are larger in size and take longer to train and evaluate.

\noindent\textbf{Layer sizes.}
The size of a layer defines the number of models in that layer. A higher number of models leads to more accurate predictions since the segments that the models have to cover are smaller.

\noindent\textbf{Error bounds.}
Error bounds facilitate the error correction by limiting the size of the interval that has to be searched. Error bounds can be chosen on different granularities or be omitted altogether.

\noindent\textbf{Search algorithm.}
Depending on the error bounds, several search algorithms may be applied to perform error correction, e.g., binary search, linear search, or exponential search.

\section{Related Work}
\label{sec:related}

The introduction of learned indexes by \ca{kraska2018case} caused both excitement and criticism within the database community. Early criticism mainly focused on the lack of efficient updates, the relatively weak baselines, and the absence of an open-source implementation~\cite{neumann2017case,bailis2018book}. Later, \ca{crotty2021histtree} claimed that the performance advantages of learned indexes are primarily due to implicit assumptions on the data such as sortedness and immutability. Subsequently published learned indexes addressed some of these weaknesses~\cite{galakatos2019fiting,ding2020alex,ferragina2020pgm}. Nevertheless, RMI remains one of the fastest indexes in experimental evaluations~\cite{kipf2019sosd,marcus2020benchmarking,kipf2020radixspline,crotty2021histtree}.

\subsection{Learned Indexes}
\label{subsec:related:learned}

Existing learned indexes commonly approximate the CDF. These indexes most notably differ in (1)~the type of model they use to approximate the CDF, (2)~whether they are trained bottom-up or top-down, and (3)~whether they support updates. \autoref{tab:related:indexes} gives an overview of existing learned indexes.

\begin{table}
    \caption{Overview of learned indexes.}
    \label{tab:related:indexes}
    \begin{footnotesize}
        \begin{tabular}{lccccc}
            \toprule
            Index & Model & Training & Updates & Open-Source \\
            \midrule
            RMI~\cite{kraska2018case}              & Multiple & top-down  & \xmark & (\cmark) \\
            FITing-tree~\cite{galakatos2019fiting} & PLA      & bottom-up & \cmark & \xmark \\
            ALEX~\cite{ding2020alex}               & PLA      & top-down  & \cmark & \cmark \\
            PGM-index~\cite{ferragina2020pgm}      & PLA      & bottom-up & \cmark & \cmark \\
            RadixSpline~\cite{kipf2020radixspline} & LS       & bottom-up & \xmark & \cmark \\
            \bottomrule
        \end{tabular}
    \end{footnotesize}
\end{table}

\noindent\textbf{FITing-tree.}
FITing-tree~\cite{galakatos2019fiting} models the CDF using \emph{piecewise linear approximation} (PLA). During training, a dataset is first divided into variable-sized segments by a greedy algorithm in a single pass over the data. The segments are created in such a way that their linear approximation satisfies a user-defined error bound. Segments are then indexed by bulk loading them into a B-tree. A lookup consists of traversing the B-tree to find the segment that contains the key, computing an estimated position based on the linear approximation of the segment, and searching the key within the error bounds around the estimated position. FITing-tree supports inserts, either in-place by shifting existing keys within the segment or using a buffering strategy, where each segment has a buffer that is merged with the other keys in the segment whenever the buffer is full. Unfortunately, at the time of writing, no open-source implementation of FITing-tree was available which kept us from including it in our experiments.

\noindent\textbf{ALEX.}
ALEX~\cite{ding2020alex} uses a variable-depth tree structure to approximate the CDF with linear models. Internal nodes are linear models which, given a key, determine the child node. Leaf nodes hold the data, the distribution of which is again approximated by a linear model. During a lookup the tree is traversed until a leaf node is reached, then a position is predicted using the leaf's linear model, and finally, the key is searched using exponential search. Like RMI, ALEX is trained top-down, however, ALEX has a dynamic structure that is controlled by a cost model, which decides how to split nodes. ALEX supports inserts by splitting or expanding full nodes.

\noindent\textbf{PGM-index.}
PGM-index~\cite{ferragina2020pgm} also approximates the CDF by means of PLA. Similar to FITing-tree, PGM-index starts by computing segments that satisfy an error bound. However, in contrast to FITing-Tree, PGM-index creates a PLA-model that is optimal in the number of segments. Each segment is represented by the smallest key in that segment and a linear function that approximates the segment. Afterwards, this process is continued recursively bottom-up by again creating a PLA-model on the smallest keys of each segment. The recursion is terminated as soon as a single segment is left. So unlike ALEX, each path from the root model to a segment is of equal length. A lookup is an iterative process where on each level of the PGM-index (1)~a linear model predicts the next-layer segment containing the key, (2)~the correct segment is searched within the error bounds around the prediction using binary search, and (3)~the process is continued for the next-layer segment until the sorted array of keys is reached. \ca{ferragina2020pgm} also introduce variants of PGM-index that support updates (dynamic PGM-index) and compression on the segment level (compressed PGM-index). The size of PGM-index depends on the number of segments required to satisfy the used-defined error bound.

\noindent\textbf{RadixSpline.}
In contrast to the aforementioned learned indexes, RadixSpline~\cite{kipf2020radixspline} approximates the CDF using a linear spline. The linear spline is fit in a single pass over the data and to satisfy a user-defined error bound. The resulting spline points are inserted into a radix table that maps keys to the smallest spline point with the same prefix. The size of the radix table depends on the user-defined prefix length. A lookup consists of finding the spline points surrounding the lookup key using the radix table, performing linear interpolation between the spline points to obtain an estimated position, and applying binary search in the error interval around the estimated position to find the key. Like RMI, RadixSpline has a fixed number of layers and does not support updates.

\subsection{Experiments and Analysis}
\label{subsec:related:eanda}

\ca{marcus2020cdfshop} published an open-source implementation of RMIs along with an automatic optimizer in December 2019. The reference implementation differs in some respects from the original description~\cite{kraska2018case}. For instance, model types like B-tree nodes and neural networks are missing and error bounds are determined on a different granularity. Given a dataset, the optimizer uses exhaustive enumeration to determine a set of pareto-optimal~(in terms of lookup time and index size) two-layer RMI configurations consisting of first-layer model type, second-layer model type, and second-layer size. Instead of blindly performing this costly enumeration, our work aims to understand the impact of each hyperparameter and to develop a simple guideline. Further, in addition to model types and layer sizes, we also consider error bounds and search algorithms when configuring RMIs.

\ca{kipf2019sosd} introduced the \emph{Search On Sorted Data}~(SOSD) benchmark, a benchmarking framework for learned indexes. Besides providing a variety of index implementations, they supply four real-world datasets. In their preliminary analysis, the authors conclude that RMI and RadixSpline are able to outperform traditional indexes including ART~\cite{leis2013adaptive}, FAST~\cite{kim2010fast}, and B-trees while being significantly smaller in size. The authors also state that the lack of efficient updates, long building times, and the need for hyperparameter tuning are notable drawbacks of learned indexes.

As a follow-up, \ca{marcus2020benchmarking} conduct a more detailed experimental analysis of learned indexes based on the framework and datasets from SOSD~\cite{kipf2019sosd}. The authors perform a series of experiments to explain the superior performance of learned indexes and conclude that a combination of fewer cache misses, branch misses, and instructions account for most of the improved performance compared to traditional indexes. Further, the authors show that learned indexes are pareto-optimal in terms of size and lookup performance independently of dataset and key size.

Both aforementioned studies~\cite{kipf2019sosd,marcus2020benchmarking} involve inventors of the RMI and aim to explain the performance of learned indexes in general. Since the evaluated RMI configurations were obtained using the optimizer~\cite{marcus2020cdfshop}, the studies neither show the impact of incorrectly configuring an RMI, nor do the studies provide advice on how to configure RMIs outside of using the optimizer. In contrast, to the best of our knowledge, we conduct the first independent and holistic analysis of RMIs that directly compares configurations and aims to explain their performance.

\ca{ferragina2020effective} take a theoretical approach at understanding the benefits of learned indexes, specifically of indexes based on PLA. The authors show that for a number of distributions, PGM-index~\cite{ferragina2020pgm}, while achieving the same query time complexity as B-trees, offers improved space complexity. To support their theoretical results, the authors conduct several experiments both on synthetic and real-world datasets. The theoretical results build a solid foundation for further research. However, since RMIs are neither limited to PLA nor do RMIs aim to construct the optimal number of segments, the results cannot be transferred to RMIs.

\section{Experimental Setup}
\label{sec:setup}

In this section, we introduce the implementation, hyperparameter settings, datasets, and workload used in our experiments and baselines considered for comparison. All experiments are conducted on a Linux machine with an Intel\textsuperscript{\textregistered} Xeon\textsuperscript{\textregistered} CPU E5-2620 v4 (2.10\,GHz, 20\,MiB L3) and 4x8\,GiB DDR4 RAM. Our code is compiled with \lstinline{clang-12.0.1}, optimization level \lstinline{-O2}, and executed single-threaded.

\subsection{Implementation}
\label{subsec:setup:impl}

Our implementation of RMIs is written in \Cpp. RMI classes have a fixed number of layers and model types are passed as template arguments. This implies that all models in a layer are of the same type. Training algorithms of the model types are adapted from the reference implementation~\cite{marcus2020cdfshop}. When assigning keys to the next-layer models, the reference implementation always copies keys to a new array. We optimized the training process based on the observation that the models considered here are monotonic and will never create overlapping segments. Thus, when assigning keys to next-layer models, we simply store iterators on the sorted array of the first and last key of each segment. We then train the next-layer models by passing them the respective iterators and thereby avoid copying the keys. Further, instead of training all models on a mapping from key to position in the sorted array, we train inner layers on a mapping from key to next-layer model index which is obtained by scaling the position to the size of the next layer similar to Equation~(\ref{eq:fi}). In other words, we train inner layers directly on a targeted equal-width segmentation. This approach saves a multiplication and division during lookup that are otherwise required for computing the model index from the position estimate. Our entire codebase is available on GitHub.\footnote{\url{https://github.com/BigDataAnalyticsGroup/analysis-rmi}}

\subsection{Hyperparameters}
\label{subsec:setup:hyper}

In the following, we give a list of hyperparameter configurations evaluated in our experiments and briefly compare them against those considered by the reference implementation's optimizer~\cite{marcus2020cdfshop}.

\begin{table*}
    \centering
    \caption{Evaluated hyperparameter configurations.}
    \label{tab:setup:hyperparameters}
    \begin{subtable}[t]{0.35\linewidth}
        \centering
        \caption{Model types}
        \label{subtab:setup:models}
        \begin{footnotesize}
            \begin{tabular}[t]{lll}
                \toprule
                Abrv. & Method & Formula \\
                \midrule
                LR & \underline{L}inear \underline{R}egression & $f(x) = ax + b$ \\
                \hline
                LS & \underline{L}inear \underline{S}pline & $f(x) = ax + b$ \\
                \hline
                CS & \underline{C}ubic \underline{S}pline & $f(x) = ax^3 + bx^2 + cx + d$ \\
                \hline
                RX & \underline{R}adi\underline{x} & $f(x) = (x \ll a) \gg b$ \\
                \bottomrule
            \end{tabular}
        \end{footnotesize}
    \end{subtable}
    \begin{subtable}[t]{0.35\linewidth}
        \centering
        \caption{Error bounds}
        \label{subtab:setup:bounds}
        \begin{footnotesize}
            \begin{tabular}[t]{lll}
                \toprule
                Abrv. & Method & Granularity \\
                \midrule
                LInd & \underline{L}ocal \underline{Ind}ividual~\cite{kraska2018case} & max +/- error per model  \\
                \hline
                LAbs & \underline{L}ocal \underline{Abs}olute~\cite{marcus2020cdfshop} & max abs error per model \\
                \hline
                GInd & \underline{G}lobal \underline{Ind}ividual & max +/- error per RMI \\
                \hline
                GAbs & \underline{G}lobal \underline{Abs}olute & max abs error per RMI \\
                \hline
                NB & \underline{N}o \underline{B}ounds~\cite{kraska2018case} & - \\
                \bottomrule
            \end{tabular}
        \end{footnotesize}
    \end{subtable}
    \begin{subtable}[t]{0.29\linewidth}
        \centering
        \caption{Search algorithms}
        \label{subtab:setup:searches}
        \begin{footnotesize}
            \begin{tabular}[t]{ll}
                \toprule
                Abrv. & Method \\
                \midrule
                Bin & \underline{Bin}ary Search \\
                \hline
                MBin & \underline{M}odel-biased \underline{Bin}ary Search~\cite{kraska2018case} \\
                \hline
                MLin & \underline{M}odel-biased \underline{Lin}ear Search \\
                \hline
                MExp & \underline{M}odel-biased \underline{Exp}onential Search~\cite{kraska2018case} \\
                \bottomrule
            \end{tabular}
        \end{footnotesize}
    \end{subtable}
\end{table*}

\noindent\textbf{Model types.}
\autoref{subtab:setup:models} lists the model types considered in our evaluation. Linear regression~(LR) is a linear model that minimizes the mean squared error (MSE). Linear spline~(LS) and cubic spline~(CS) fit a linear respectively cubic spline segment through the leftmost and rightmost data points. Radix~(RX) eliminates the common prefix and maps keys to their most significant bits. Models most notably differ in three respects.

\begin{enumerate}[leftmargin=*]
    \item \emph{Built time.}
        LS, CS, and RX are fast to build from the leftmost and rightmost key. LR, a regression method, is built on all keys.
    \item \emph{Evaluation time.}
        RX is the fastest to evaluate with only two bit shifts. LR and LS are equally fast to evaluate, CS is the slowest.
    \item \emph{Predictive quality.}
        LS and CS are spline techniques whose predictive quality is based on how representative the leftmost and rightmost keys are. LR minimizes the error across all keys. RX is radix-based and therefore only used for segmentation.
\end{enumerate}
In addition to the four models listed, the optimizer~\cite{marcus2020cdfshop} considers radix tables and a specialized variant of linear regression (see \autoref{subsec:comparison:lookup}) for the first layer and cubic splines for the second layer. We decided to evaluate a smaller set of model types to analyze the impact of model types in general. Since the optimizer always recommends LR for the second layer, we only consider LR and LS for the second layer.

\noindent\textbf{Layer count.}
Like the optimizer~\cite{marcus2020cdfshop}, we only consider two-layer RMIs. It was previously reported that in most cases two layers are sufficient to accurately approximate a CDF~\cite{marcus2020cdfshop,marcus2020benchmarking}, which we verified for the considered datasets in preliminary experiments. We plan to explore multi-layer RMIs as part of future work.

\noindent\textbf{Layer size.}
We cover the same wide range of second layer sizes between~$2^6$ and~$2^{25}$ in power of two steps like the optimizer~\cite{marcus2020cdfshop}.

\noindent\textbf{Error bounds.}
We consider five different variants of error bounds listed in \autoref{subtab:setup:bounds}, which differ in the granularity of the stored bounds in two respects. (1)~Error bounds might either be computed for each last-layer model (local) or for the entire RMI (global). Global bounds, while being more memory efficient, are prone to outliers as the single largest error determines the search interval size of all lookups. Local bounds are more robust against outliers as an outlier only affects the respective model. (2)~We can either store the maximum absolute error (absolute) or both the maximum positive and negative error individually (individual). While the former is again more space efficient, the latter allows for tighter bounds, especially, if a model either overestimates or underestimates the actual position.
Additionally, we might not store any bounds (NB). Both local individual (LInd) and NB were suggested by \ca{kraska2018case}. The reference implementation supports local absolute bounds (LAbs) and NB, but the optimizer~\cite{marcus2020cdfshop} always recommends LAbs.

\noindent\textbf{Search algorithm.}
The evaluated search algorithms are listed in \autoref{subtab:setup:searches}. We generally distinguish between two types of search algorithms: (1)~search algorithms that only consider the error bounds and (2)~search algorithms that also utilize the estimated position (model-biased)~\cite{kraska2018case}. Standard binary search is an example of the first type of search algorithm. We search the key in the interval between the two error bounds and ignore the position estimate. However, binary search can be adjusted to become model-biased. Instead of choosing the middle element of the interval as first comparison point, we pick the estimated position. Similarly, linear search and exponential search can be tweaked to become model-biased. Instead of searching the interval from left to right, we start the search from the estimated position and search to the left or right, depending on whether the prediction is an overestimation or an underestimation. The search is stopped once it is certain that the key cannot be found anymore. Initially, we also considered standard linear search and exponential search for our experiments but both always performed worse than their model-biased counterparts. Note that not all combinations of error bounds and search algorithms make sense, e.g.,~in the case of absolute error bounds, model-biased binary search and standard binary search are essentially the same as the estimate will be the center of the interval. Further, model-biased linear and exponential search do not require bounds. Previous studies compared binary~\cite{kraska2018case,kipf2019sosd,marcus2020benchmarking}, linear, and interpolation search~\cite{marcus2020benchmarking}. Model-biased variants of linear and exponential search have not been studied in the context of RMIs so far.

\subsection{Datasets}
\label{subsec:setup:datasets}

\begin{figure}[b]
    \includegraphics[scale=0.4]{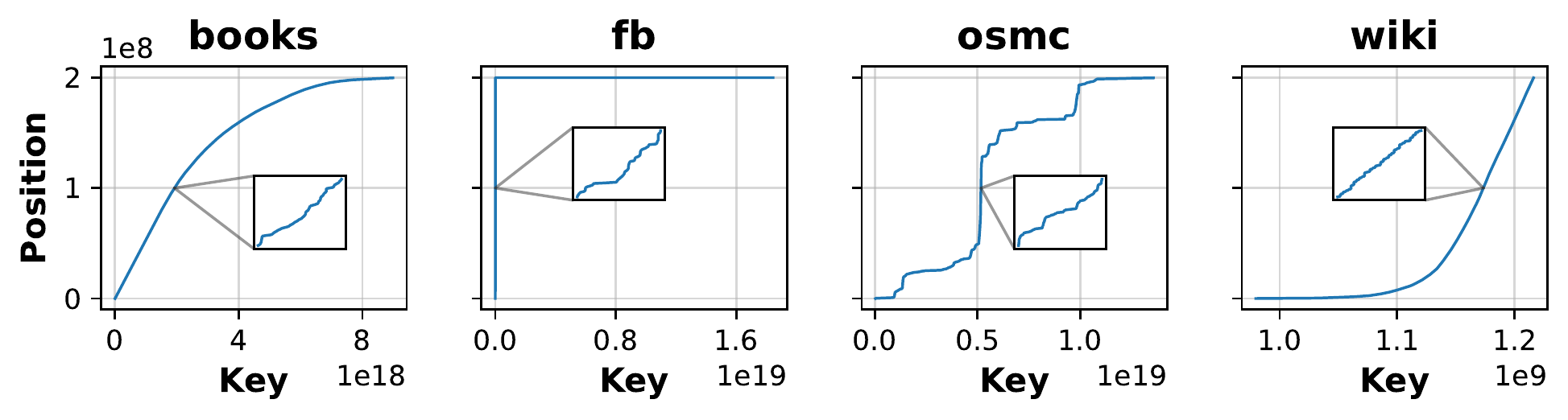}
    \caption{CDFs of four real-world datasets from SOSD~\cite{kipf2019sosd}. Zoom-ins show segments of 100 consecutive keys.}
    \label{fig:setup:datasets:cdf}
\end{figure}

Learned indexes are know to adapt well to artificial data sampled from statistical distributions~\cite{marcus2020benchmarking}. Therefore, we use the four real-world datasets from the SOSD benchmark~\cite{kipf2019sosd}. Each dataset consists of 200M 64-bit unsigned integer keys. The CDFs of the four datasets are depicted in \autoref{fig:setup:datasets:cdf}, zoom-ins show a segment of 100 consecutive keys and indicate the amount of noise in the dataset.

\noindent\textbf{\books}: keys represent the popularity of books on Amazon.

\noindent\textbf{\fb}: keys represent Facebook user ID. This dataset contains a small number of extreme outliers, which are several orders of magnitude larger than the rest of the keys, at the upper end of the key space. These outliers were not plotted in previous studies~\cite{kipf2019sosd,marcus2020benchmarking}.

\noindent\textbf{\osmc}: keys represent cell IDs on OpenStreetMap. This dataset has clusters that are artifacts of projecting two-dimensional data into one-dimensional space~\cite{marcus2020benchmarking}.

\noindent\textbf{\wiki}: keys are edit timestamps on Wikipedia, contains many duplicates.

\subsection{Workload}
\label{subsec:setup:workload}

For the lookup performance, we consider lower bound queries, i.e.,~for a given key, the index returns an iterator to the smallest element in the sorted array that is equal to or greater than the key. The sorted array is kept in memory and we perform 20M lookups per run, the keys of which are sampled from the sorted array uniformly at random with a fixed seed. Reported execution times are the average execution time of the median of three runs.

\subsection{Baselines}
\label{subsec:setup:baselines}

In \autoref{sec:comparison}, we compare our RMI implementation against a number of baselines listed in \autoref{tab:setup:indexes} for which we use the referenced open-source implementations. Due to our focus on ranking the performance of RMIs, we consider all publicly available learned indexes but only some representatives of traditional indexes.

\begin{table}
    \centering
    \caption{Overview of the considered baselines.}
    \label{tab:setup:indexes}
    \begin{footnotesize}
        \begin{tabular}{lllc}
            \toprule
            Method & Type & Hyperparameters & Source \\
            \midrule
            RMI~\cite{kraska2018case} & Learned & model types, layer size & \cite{rmicode} \\
            ALEX~\cite{ding2020alex} & Learned & sparsity & \cite{alexcode} \\
            PGM-index \cite{ferragina2020pgm} & Learned & max error & \cite{pgmcode} \\
            RadixSpline \cite{kipf2020radixspline} & Learned & radix width, max error & \cite{radixsplinecode} \\
            \hline
            B-tree \cite{bayer72organisation} & Tree & sparsity & \cite{tlxcode} \\
            Hist-Tree \cite{crotty2021histtree} & Tree & num bins, max error & \cite{histtreecode} \\
            \hline
            ART \cite{leis2013adaptive} & Trie & sparsity & \cite{sosdcode} \\
            \hline
            Binary search & Search & - & \cite{cppreflowerbound} \\
            \bottomrule
        \end{tabular}
    \end{footnotesize}
\end{table}

\noindent\textbf{Learned indexes.}
ALEX~\cite{ding2020alex}, PGM-index~\cite{ferragina2020pgm}, and RadixSpline~\cite{kipf2020radixspline} are learned indexes discussed in \autoref{subsec:related:learned}. The index size of PGM-index and RadixSpline is varied based on the maximum error. Additionally, RadixSpline provides a parameter to adjust the size of the radix table that is used to index the spline points. Since we do not consider update performance here, we use the standard variant of PGM-index, which does not support updates. ALEX does not provide any parameters itself, so we vary its size by adjusting the number indexed keys (sparsity) by inserting only every $k$-th key. In addition, we also consider the reference implementation of RMIs~\cite{marcus2020cdfshop} that is configured using its integrated optimizer.

\noindent\textbf{Traditional indexes.}
B-tree~\cite{bayer72organisation} and ART~\cite{leis2013adaptive} are traditional in-memory index structures. We vary the size of B-tree and ART by adjusting the number of keys that are inserted. Therefore, we use an implementation of ART that supports lower bound queries from SOSD~\cite{kipf2019sosd}. The recently published Hist-Tree~\cite{crotty2021histtree} is a tree-structured index. Each inner node in a Hist-Tree is a histogram that partitions the data into equal-width bins. Like learned indexes, Hist-Tree exploits that the data is sorted. Hist-Tree provides two tuning parameters: the number of bins determines the size of inner nodes and the maximum error defines a threshold for the size of a terminal node. We use an implementation of a Compact Hist-Tree that does not support updates in favor of lookup performance~\cite{histtreecode}.

\noindent\textbf{Binary search.}
We also consider standard binary search over the sorted array without any index as provided by \lstinline{std::lower_bound}.

\section{Predictive Accuracy Analysis}
\label{sec:error}

In this section, we analyze the impact of hyperparameters on the predictive accuracy of RMIs. Our analysis is divided into three parts. \\
\noindent\textbf{Segmentation} (\autoref{subsec:error:segmenting}): We investigate how root models of different types divide the keys into segments. \\
\noindent\textbf{Position Prediction} (\autoref{subsec:error:pred}): We analyze how accurately different combinations of models approximate the CDF. \\
\noindent\textbf{Error Bounds} (\autoref{subsec:error:bounds}): We examine how different types of error bounds limit the error interval to be searched.

\subsection{Segmentation}
\label{subsec:error:segmenting}

\begin{figure}
    \centering
    \includegraphics[scale=0.4]{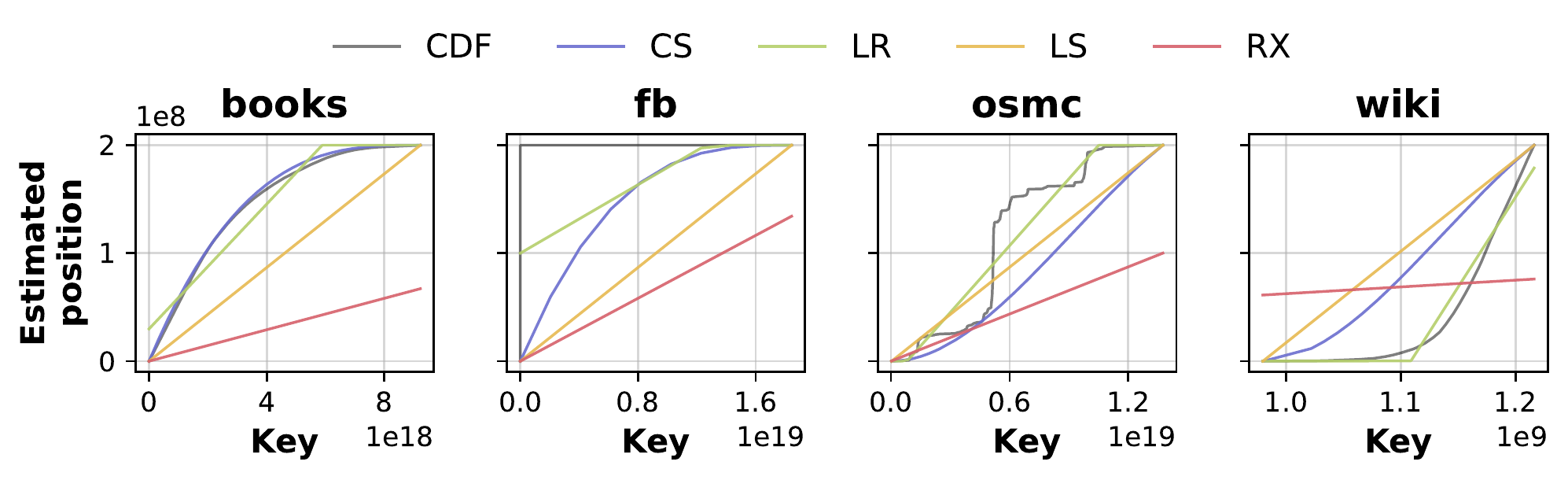}
    \caption{CDF approximations by root models of different types based on which keys are segmented.}
    \label{fig:error:single-layer}
\end{figure}

An RMI divides the keys into segments based on its root model's approximation of the CDF. Assuming a root model correctly predicts the position of each key, each segment would consist of the same number of keys. Therefore, RMIs aim for an equal-depth segmentation by design. This approach to segmentation has a crucial weakness: it ignores whether the resulting segments can be accurately approximated by the next-layer models. In contrast, other learned indexes like PGM-index~\cite{ferragina2020pgm} and RadixSpline~\cite{kipf2020radixspline}, which are built bottom-up, explicitly create segments that meet a certain error tolerance. Consequently, the quality of an RMI's segmentation cannot be assessed independently of the next layer. In the following, we address two problems that may occur when segmenting keys in an RMI: (1)~\emph{empty segments}, which do not contain any keys, and (2)~\emph{large segments}, which contain significantly more keys than others. \autoref{fig:error:single-layer} shows the CDFs and the corresponding root model approximations.

\begin{figure}
    \centering
    \includegraphics[scale=0.4]{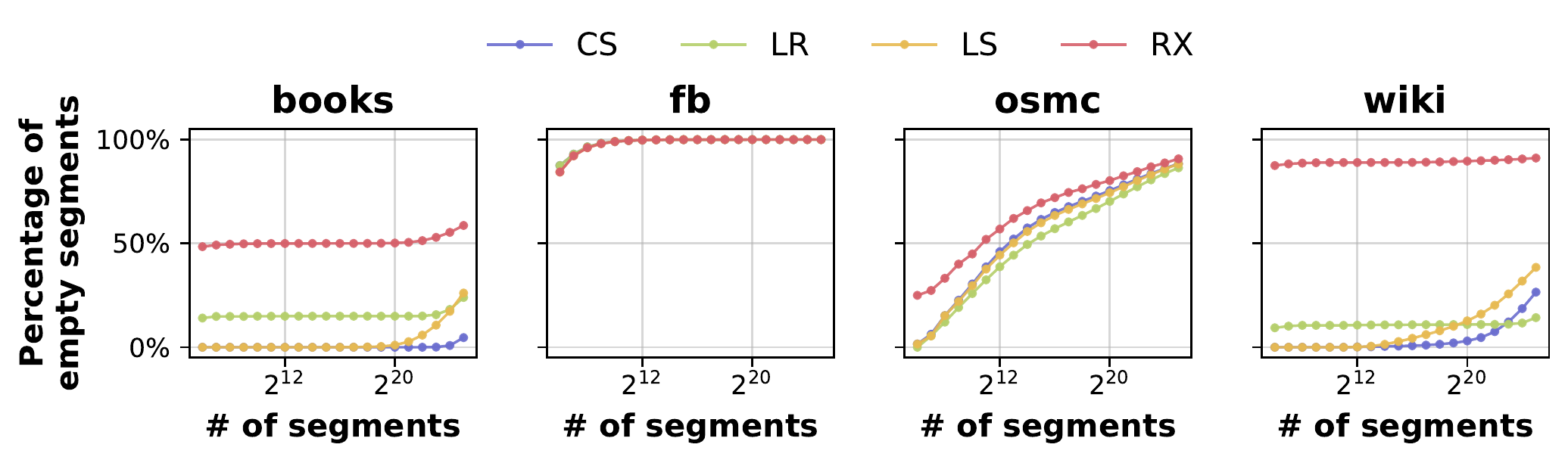}
    \caption{Percentage of empty segments when segmenting the keys with different first-layer models.}
    \label{fig:error:empty-segments}
\end{figure}

\noindent\textbf{Empty segments.}
Since there is a second-layer model for every segment, empty segments increase the size of an RMI without improving the prediction accuracy. Thus, we should aim for as few empty segments as possible. \autoref{fig:error:empty-segments} shows the percentage of empty segments of each model type on each dataset for a varying number of segments. We generally observe that the percentage of empty segments increases with an increasing number of segments. The more accurately a model approximates the CDF, the fewer empty segments it creates. For instance, CS produces empty segments on \books only after a high number of segments is reached. In contrast, radix predictions often do not cover the full range of positions, e.g.,~on \wiki, leaving the segments associated with the non-covered positions empty. The clustered distribution of \osmc dataset causes percentages to be generally higher and to increase more quickly since the keys are distributed over a small number of segments. Due to the few extreme outliers that strongly affect the CDF approximation of \fb, all models map the majority of keys to the same position, causing all of these keys to be assigned to the same segment. Increasing the number of segments gradually removes the outliers from this segment, but the segment will continue to contain most keys.

\begin{figure}
    \centering
    \includegraphics[scale=0.4]{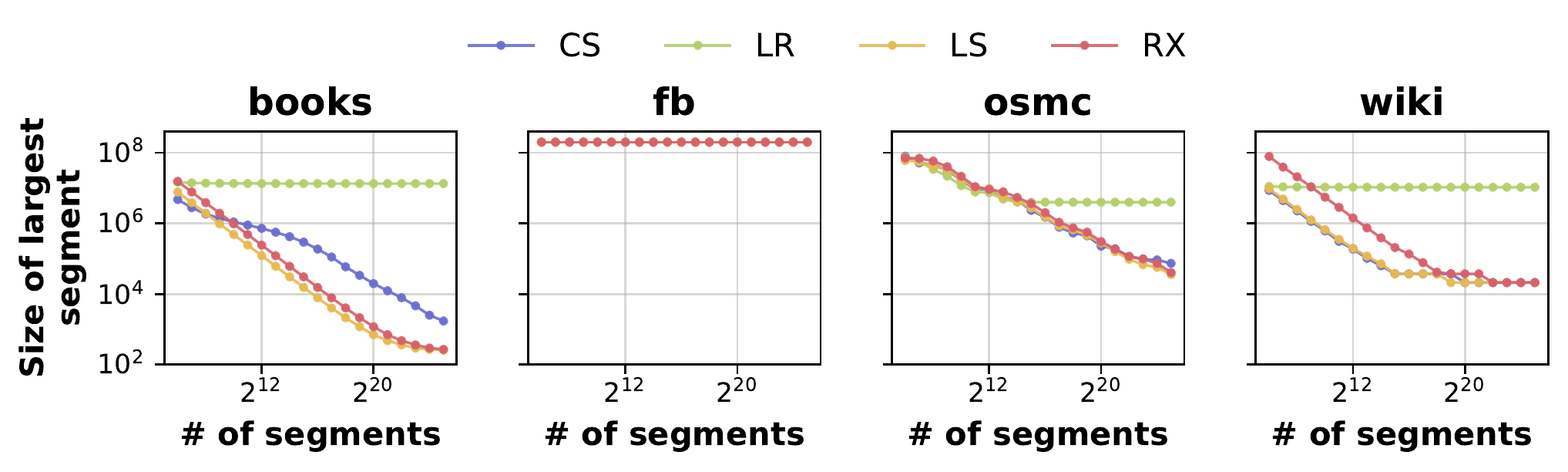}
    \caption{Size of the largest segment when segmenting the keys with different first-layer models.}
    \label{fig:error:largest-segment}
\end{figure}

\noindent\textbf{Large segments.}
Large segments potentially follow a more complex distribution and are more difficult to approximate by the second-layer models. Therefore, large partitions may negatively affect the prediction quality of an RMI. \autoref{fig:error:largest-segment} shows the number of keys that reside in the largest segment. Again, the more accurate a model approximates the CDF, the more evenly the keys are distributed over the segments. Logically, the average segment size decreases as the number of segments increases. However, this does not necessarily apply to the largest segment. For LR, the size of the largest partition often remains near-constant. The reason for this is that LR may produce estimates outside the range of valid positions. These out-of-range predictions are then clamped to either the first or last valid position. All keys whose prediction is clamped will be assigned to the same segment. Increasing the number of segments only decreases the size of these segments until the segments consist exclusively of keys whose prediction had to be clamped. CS, LS, and RX do not produce estimates outside the range of valid positions and therefore do not exhibit this problem. As discussed before, on \fb, almost all keys reside in a single segment, regardless of the number of segments and type of the root models. As we will see in subsequent experiments, the inability of the considered model types to segment datasets with extreme outliers is the main reason for inaccurate predictions, large error intervals, and slow lookups on \fb.

\noindent\textbf{Summary.}
When choosing a first-layer model type for segmentation, empty and large segments should be avoided. In our experiments, LS and CS produced the most uniform segments. RX tends to produce many empty segments. LR often creates large segments at the upper and lower end of the key space due to clamping. If none of the models satisfactorily segments the keys as with \fb, more complex models must be considered.

\subsection{Position Prediction}
\label{subsec:error:pred}

\begin{figure}
    \centering
    \includegraphics[scale=0.4]{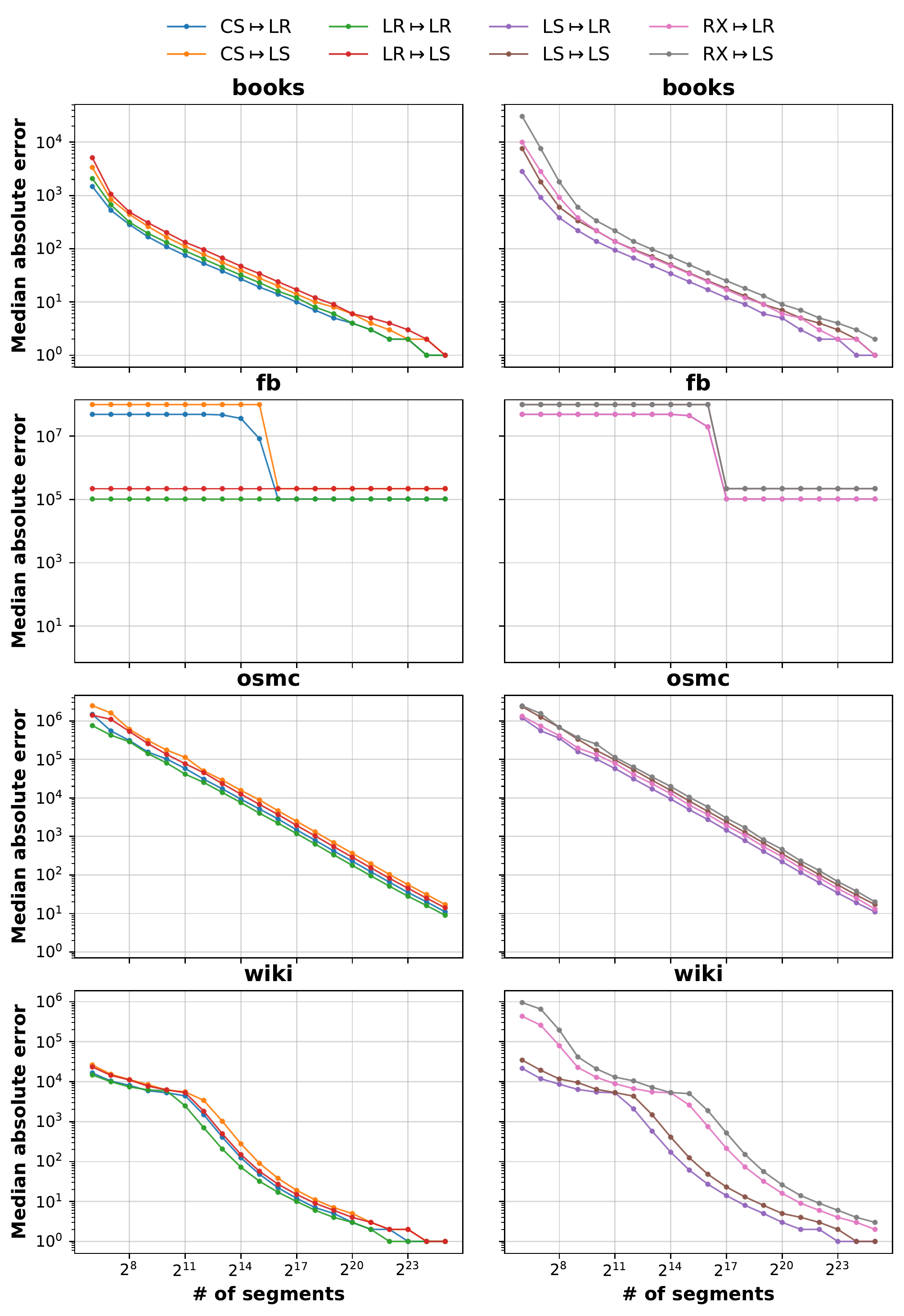}
    \caption{Median absolute error of RMIs with different combinations of first-layer and second-layer models.}
    \label{fig:error:median-ae}
\end{figure}

To analyze the impact of model types on prediction accuracy, we train RMIs of all combinations of first-layer and second-layer model types with different second-layer sizes on the four datasets. In \autoref{fig:error:median-ae}, we report the median absolute error over all keys as a measure of deviation between predicted position and actual position. We decided against reporting the mean absolute error due to variances caused by high errors on the large partitions when segmenting with LR. In the remainder, we refer to an RMI that uses RX and LR in the first and second layer, respectively, as RX$\mapsto$LR.

As expected, RMIs with more segments and thus more second-layer models generally produce more accurate predictions. On both the \books and \wiki dataset, RMIs with more than $2^{19}$ second-layer models even achieve errors in single digits. The \osmc and \fb dataset are more difficult to approximate. The \osmc dataset has a clustered distribution that results in a high number of empty segments, making non-empty segments larger on average. Additionally, these segments often have a significant amount of noise and cannot be approximated precisely with the models considered here. Similarly, the large prediction error of \fb can also be attributed to the single large segment. The sudden drop in prediction error between $2^{15}$ and $2^{17}$ segments is due to fewer of the outliers being assigned to the large segment anymore. Although the distribution within that large segment is close to uniform, it still contains a considerable amount of noise that leads to the persistent high prediction error.

Comparing the different RMI configurations, RMIs with LR, LS, and CS as root model achieve similar errors while RX performs slightly worse. This indicates that in terms of prediction accuracy, RX is less suitable for segmentation. Regarding the second-layer models, LR always achieves lower errors than LS. This is expected since LR is the only regression model and minimizes the MSE.

\noindent\textbf{Summary.}
For the first layer, a segmentation that distributes the keys over many models is a prerequisite for high prediction accuracy. For the second layer, regression models like LR achieve higher accuracy than spline models since regression models minimize the prediction error. Increasing the second-layer size of an RMI further improves its accuracy. Overall, LS$\mapsto$LR and CS$\mapsto$LR achieve good accuracy across datasets, except for \fb due to poor segmentation.

\subsection{Error Bounds}
\label{subsec:error:bounds}

\begin{figure}
    \centering
    \includegraphics[scale=0.4]{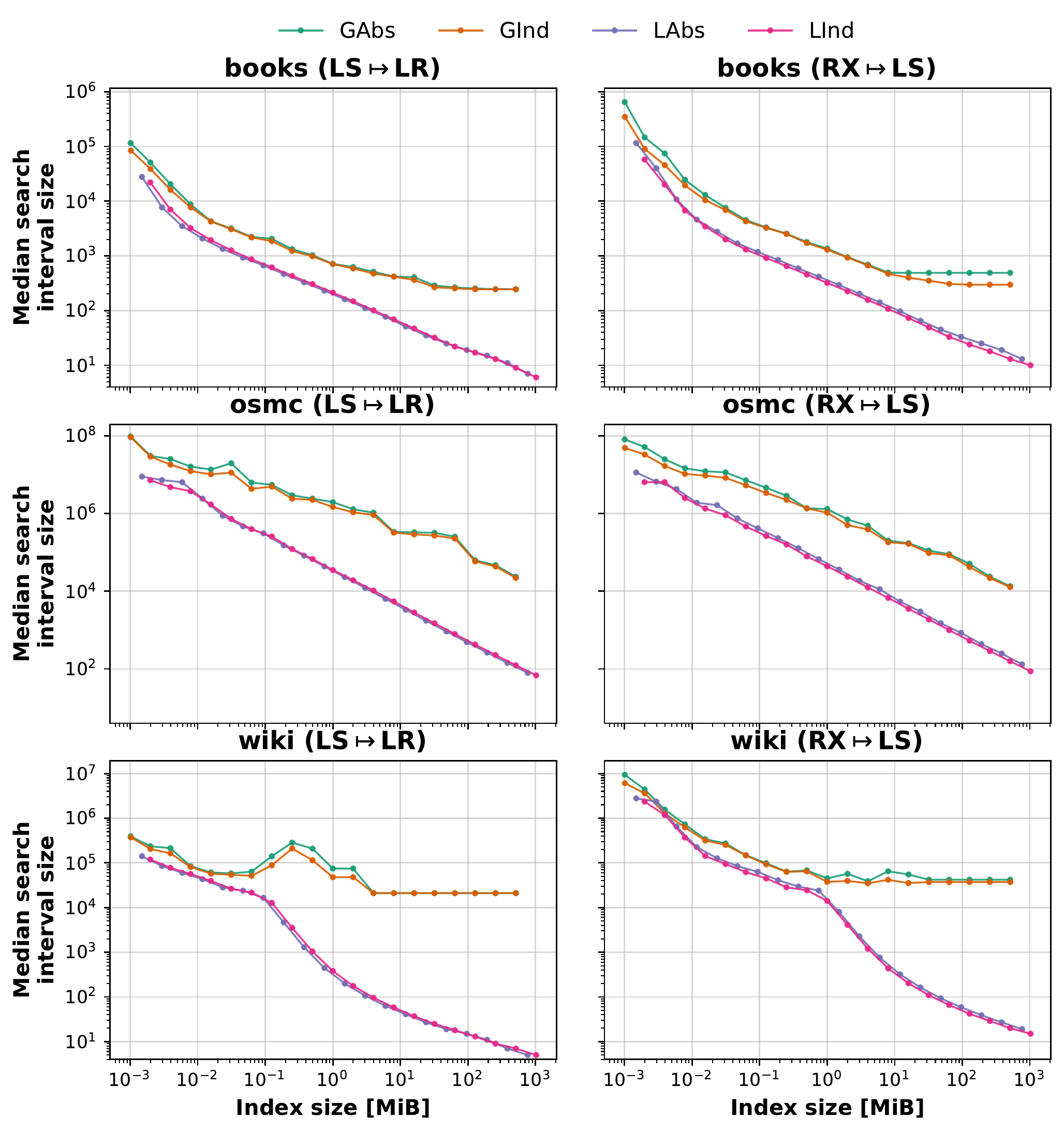}
    \caption{Comparison of error interval sizes for different error bounds for two example model  combinations.}
    \label{fig:error:median-interval}
\end{figure}

Error bounds facilitate correcting prediction errors by limiting the size of interval that has to be searched during a lookup. To evaluate the impact of different error bounds, we again train RMIs with all combinations of first-layer and second-layer model type and varying second-layer size. For each configuration, we compute error bounds of different types and record the error interval sizes over all keys. In \autoref{fig:error:median-interval}, we report the median error interval size, i.e.,~the median number of keys that have to be searched during a lookup. Here, we only show two combinations of models and omit \fb as the size of the error interval remains near constant due to inaccurate predictions. The complete experimental results can be found in \autoref{fig:appendix:error:median-interval} in \autoref{sec:results}.

Global bounds consistently lead to significantly larger error intervals than local bounds, despite the fact that at a similar index size, global bounds allow for more second-layer models and achieve on average more accurate predictions. Global bounds, however, are prone to single bad predictions, whereas local bounds are more robust because they refer to only one model. LInd and LAbs achieve similar error interval sizes. Spline models, which tend to either overestimate or underestimate, profit from LInd. LR, which often achieves similar positive and negative errors, works better with LAbs as LAbs allows for more second-layer models at a similar size.

\noindent\textbf{Summary.}
Considering RMIs of similar size, local bounds consistently result in smaller error intervals than global bounds. For the preferred second-layer model type LR, LAbs achieves smaller error intervals due to more second-layer models at a similar index size.

\section{Lookup Time Analysis}
\label{sec:lookup}

In this section, we analyze the impact of hyperparameters on the lookup performance of RMIs. Our analysis is divided into two parts. \\
\noindent\textbf{Model Types} (\autoref{subsec:lookup:model}): We investigate the lookup performance of different combinations of first and second-layer model types. \\
\noindent\textbf{Error Correction} (\autoref{subsec:lookup:error_correction}): We analyze the impact of error bounds and search algorithms on lookup performance.

\subsection{Model Types}
\label{subsec:lookup:model}

\begin{figure}
    \centering
    \includegraphics[scale=0.4]{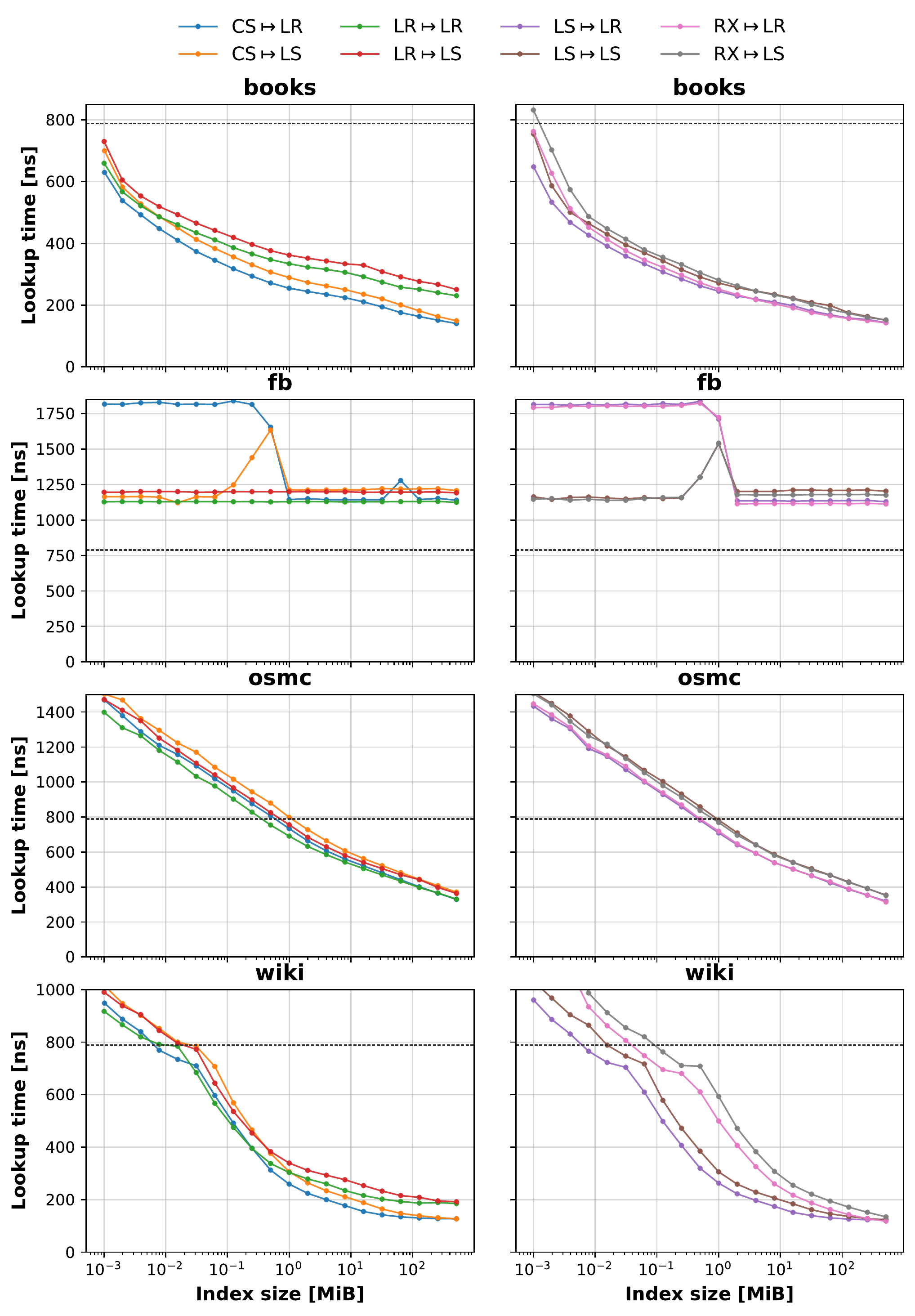}
    \caption{Comparison of lookup time for different combinations of models with NB+MExp for error correction.}
    \label{fig:lookup:models}
\end{figure}

To evaluate the impact of model types on lookup performance, we train RMIs of all combinations of first-layer and second-layer model type with varying second-layer sizes. We use no bounds and model-biased exponential search (NB+MExp) for error correction as this configuration relies solely on the predictive power of the RMI and thus most clearly illustrates the differences between the various combinations of model types. In \autoref{fig:lookup:models}, we report the average lookup time of each configuration. The dashed horizontal lines are the average time for obtaining a key using binary search.

For a fixed index size, the lookup times of different models within a dataset often differ only slightly, e.g.,~on \osmc and \books, all combinations of models have similar lookup times. However, lookup times vary significantly across different datasets. This observation is consistent with the prediction errors we saw in \autoref{subsec:error:pred}. The reason for this is that lookup time consists of evaluation time and error correction time. The error correction time accounts for the majority of lookup time and is determined by the prediction error. However, balancing evaluation time and error correction time is a trade-off that has to be carefully considered. In our experiments, we only consider relatively simple models that are fast to evaluate and, as a result, there are only minor differences in evaluation time. In preliminary experiments, we also considered neural networks, which achieved higher prediction accuracy, but the faster error correction was overshadowed by a significantly higher evaluation time ultimately resulting in considerably slower lookups. Of the models considered here, CS is the slowest to evaluate. We can observe the impact of its slower evaluation time compared to LS on \books where despite CS$\mapsto$LR being slightly more accurate than LS$\mapsto$LR, LS$\mapsto$LR achieves faster lookups. Differences in evaluation time are particularly noticeable when the error correction time is relatively short which often is the case for configurations that are larger in size.

\noindent\textbf{Summary.}
Prediction accuracy is a strong indicator for lookup performance as it determines the error correction time. Therefore, models like CS$\mapsto$LR and LS$\mapsto$LR that achieve good accuracy across datasets should be chosen. However, the more accurate the predictions are, the more important become differences in evaluation time and models that are slightly less accurate but faster to evaluate have an advantage. Increasing the second-layer size improves accuracy and causes the lookup time to converge.

\subsection{Error Correction}
\label{subsec:lookup:error_correction}

\begin{figure}
    \centering
    \includegraphics[scale=0.4]{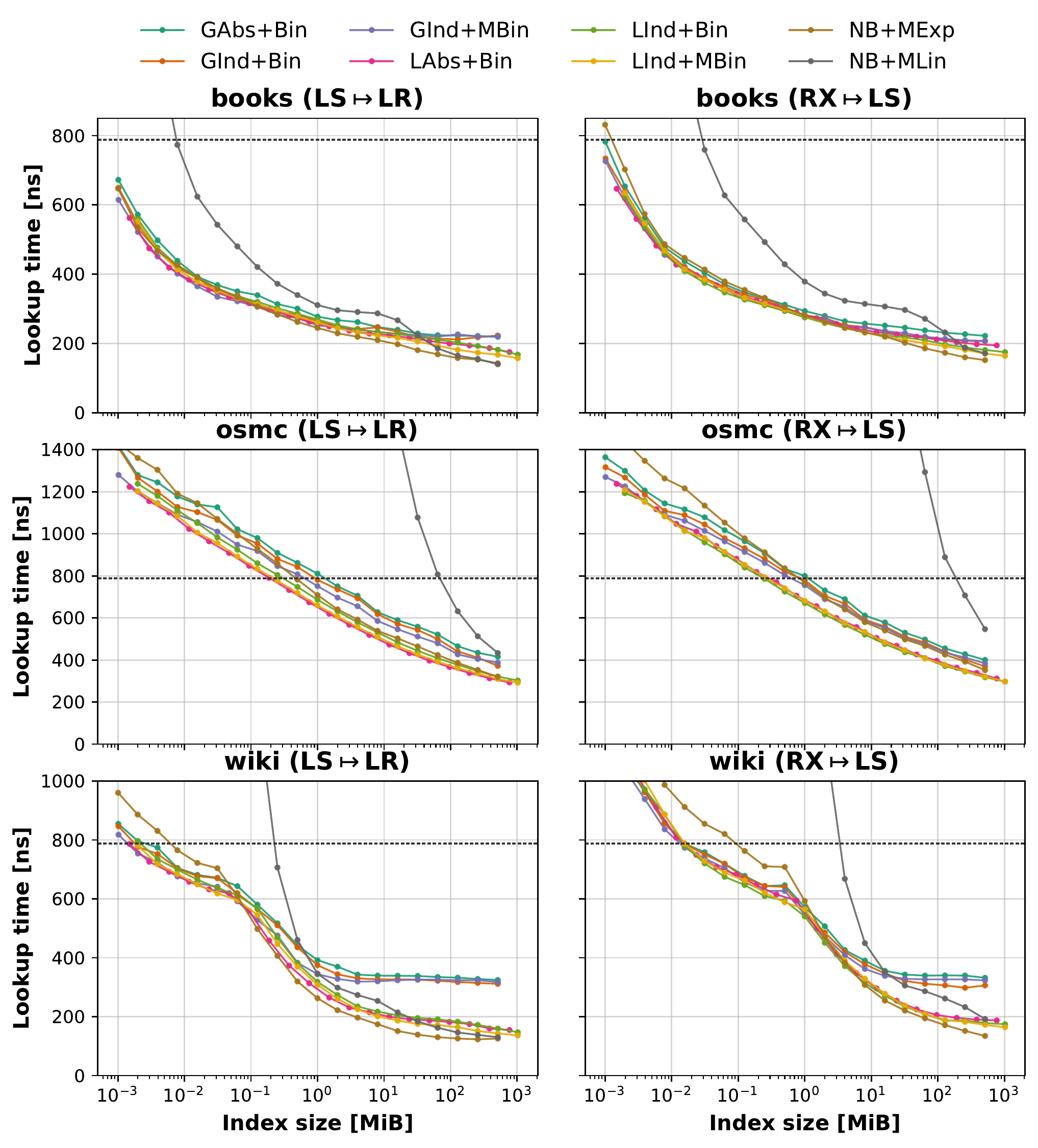}
    \caption{Comparison of lookup time for different error corrections using two combinations of models as examples.}
    \label{fig:lookup:error-correction}
\end{figure}

Next, we examine the impact of eight combinations of error bounds and search algorithms for error correction on lookup time. We consider the following combinations. NB is evaluated with MLin and MExp as both search algorithms do not use bounds. GInd and LInd are evaluated with MBin and Bin. GAbs and LAbs are evaluated with Bin only as MBin and Bin are the same in case of absolute bounds. In \autoref{fig:lookup:error-correction}, we report the average lookup time. Here, we again show only two combinations of models and omit \fb because lookup performance could not be significantly improved compared to \autoref{fig:lookup:models} in this experiment. The complete experimental results can be found in \autoref{fig:appendix:lookup:error-correction} in \autoref{sec:results}.

We observe that either a configuration with local bounds or without any bounds performs best. Local bounds generally perform better than global bounds, which is consistent with our observation from \autoref{subsec:error:bounds}. Nevertheless, binary search mitigates differences in search interval size drastically, e.g.,~global and local bounds perform almost identical with LS$\mapsto$LR on \books, although the search interval sizes differ by more than an order of magnitude. LInd and LAbs perform almost identical with a maximum performance difference of factor 1.1x. Similar to what we saw in \autoref{subsec:error:bounds}, LS works better with LInd as it tends to either overestimate or underestimate, LR works better with LAbs as its loss function often causes the maximum overestimation and underestimation to be similar. Considering LInd, there is hardly any difference between Bin and MBin.

Similar to what we saw in \autoref{subsec:lookup:model} with respect to model types, the choice of error bounds not only affects error correction time but also evaluation time as error bounds induce overhead for computing the search interval's limits. Hence, RMIs without error bounds are faster to evaluate. The faster evaluation is particularly noticeable when the RMI achieves a high prediction accuracy and thus fast error correction. In these cases, NB+MExp performs better than configurations with bounds as can be seen with \books and \wiki. To further analyze when to use NB+MExp over configurations with bounds, we also recorded the mean log$_2$ error as an estimate of the number of search steps required by MExp. Starting at an mean log$_2$ error of around 7 to 10, NB+MExp is faster than LAbs+Bin. NB+MLin requires even lower errors to be similarly fast as NB+MExp.

\noindent\textbf{Summary.}
The best combination of error bounds and search algorithm depends on the predictive accuracy of the RMI. If the mean log$_2$ error is sufficiently small, NB+MExp performs best due to RMIs without bounds being faster to evaluate. For larger errors, configuration with local bounds such as LAbs+Bin perform better.

\section{Build Time Analysis}
\label{sec:build}

\begin{figure}
    \centering
    \begin{subfigure}{.3\linewidth}
        \centering
        \includegraphics[scale=0.4]{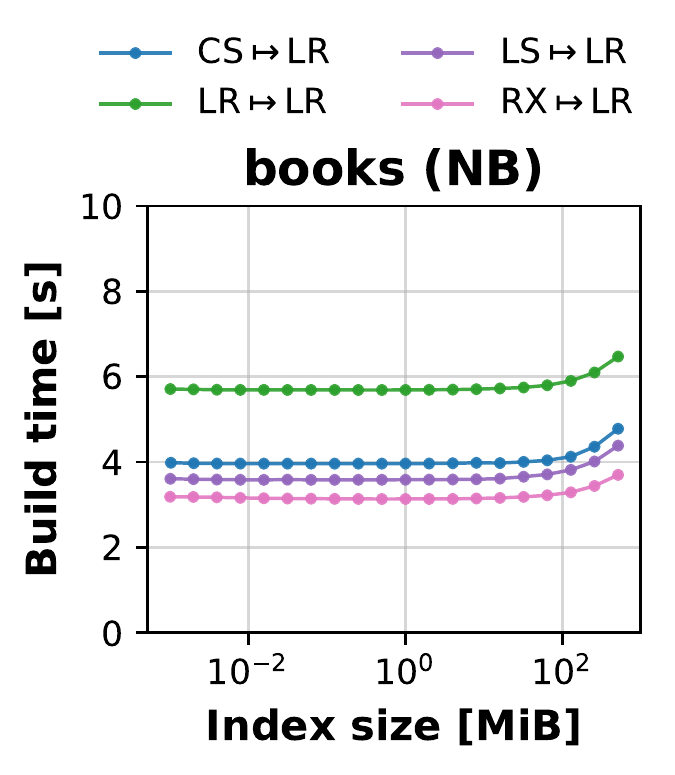}
        \caption{Layer 1 type}
        \label{subfig:build:ours:layer1}
    \end{subfigure}
    \begin{subfigure}{.3\linewidth}
        \centering
        \includegraphics[scale=0.4]{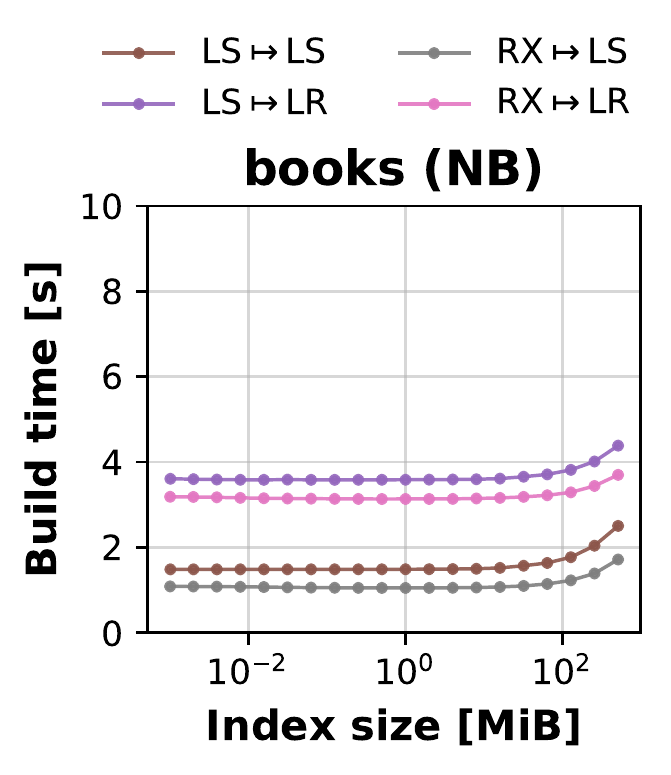}
        \caption{Layer 2 type}
        \label{subfig:build:ours:layer2}
    \end{subfigure}
    \begin{subfigure}{.3\linewidth}
        \centering
        \includegraphics[scale=0.4]{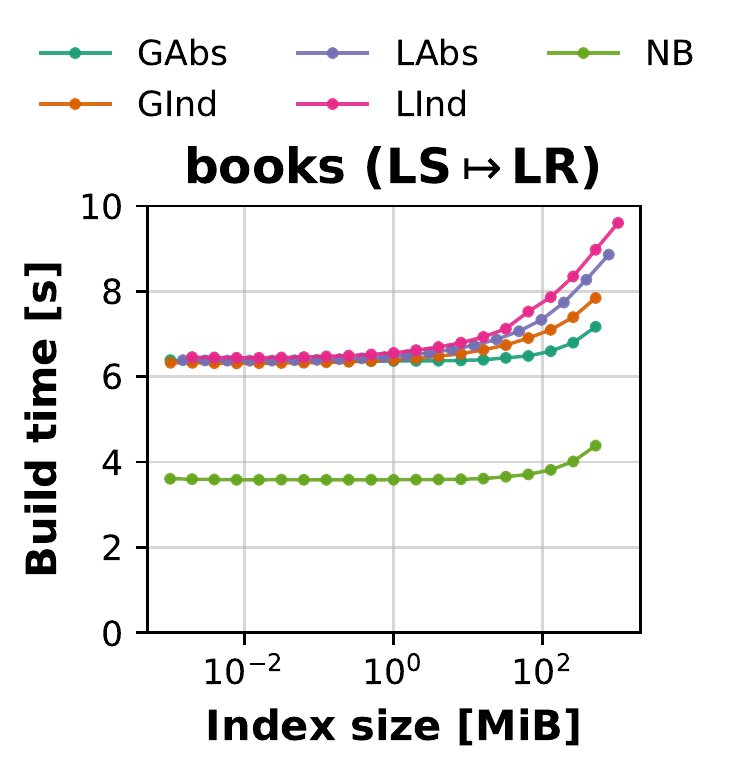}
        \caption{Error bound}
        \label{subfig:build:ours:bounds}
    \end{subfigure}
    \caption{Build times when varying hyperparameters.}
    \label{fig:build:ours}
\end{figure}

In this section, we analyze the build time of our implementation of RMIs and compare it with the reference implementation~\cite{marcus2020cdfshop}. Recall that the build process of a two-layer RMI consists of four steps: (1)~training the root model, (2)~creating segments, (3)~training the second-layer models, and (4)~computing error bounds. \autoref{fig:build:ours} shows build times on \books. Except for minor caching effects on large configurations, the build time is independent of the dataset. The complete experimental results can be found in \autoref{fig:appendix:build:ours:bounds} in \autoref{sec:results}.
We discuss each aspect that affects build time individually below.

\noindent\textbf{First-layer type.}
Consider \autoref{subfig:build:ours:layer1} for a build time comparison of different root models. Models in general and root models in particular not only differ in training time, which affects step (1), but also in evaluation time, which affects steps (2) and (4). The most notable difference between the models in terms of training time is whether a model considers all keys, like LR, or a constant number of keys, like LS, CS, and RX. Since the evaluation time of LR and LS is the same, the difference in build time in \autoref{subfig:build:ours:layer1} can be attributed entirely to the training time of the root model. Like LS, RX also considers only two keys for training. Here, the faster build time of RX is caused by the faster evaluation of RX during segmentation. CS is faster than LR because it again only considers a constant number of keys but slower than LS because training and evaluation are slightly slower.

\noindent\textbf{Second-layer type.}
Consider \autoref{subfig:build:ours:layer2} for a build time comparison of different second-layer models. Analogously to the root model type, the second-layer model type affects training time and evaluation time. Second layers consisting of LS models takes about two seconds less to train than second layers consisting of LR models. Note that in this example, the second layer is never evaluated because we do not compute bounds. Otherwise, evaluation time would be the same for LR and LS.

\noindent\textbf{Error bounds.}
Consider \autoref{subfig:build:ours:bounds} for a build time comparison of different error bounds. Computing error bounds requires evaluating the RMI on every key plus the actual computation of the bounds. This additional effort explains the difference in built time between NB and configurations with bounds. The difference between individual configurations with bounds is mainly due to branch misses when calculating the bounds. At similar index size, local bounds trigger more branch misses than global bounds and individual bounds trigger more branch misses than absolute bounds.

\noindent\textbf{Index size.}
Consider again the RMI configuration without bounds in \autoref{subfig:build:ours:bounds}. The build time remains almost constant as long as the entire RMI fits in cache (20\,MiB). Once the RMI no longer fits in cache, the build time increases due to cache misses. Next, consider the configurations with bounds in \autoref{subfig:build:ours:bounds}. Here, the previously described branch and cache misses add up and the build time already increases for configurations that are smaller than the cache size. The increase in build time is less pronounced if a configuration produces many empty segments due to less cache misses.

\begin{figure}
    \centering
    \begin{subfigure}{.45\linewidth}
        \centering
        \includegraphics[scale=0.4]{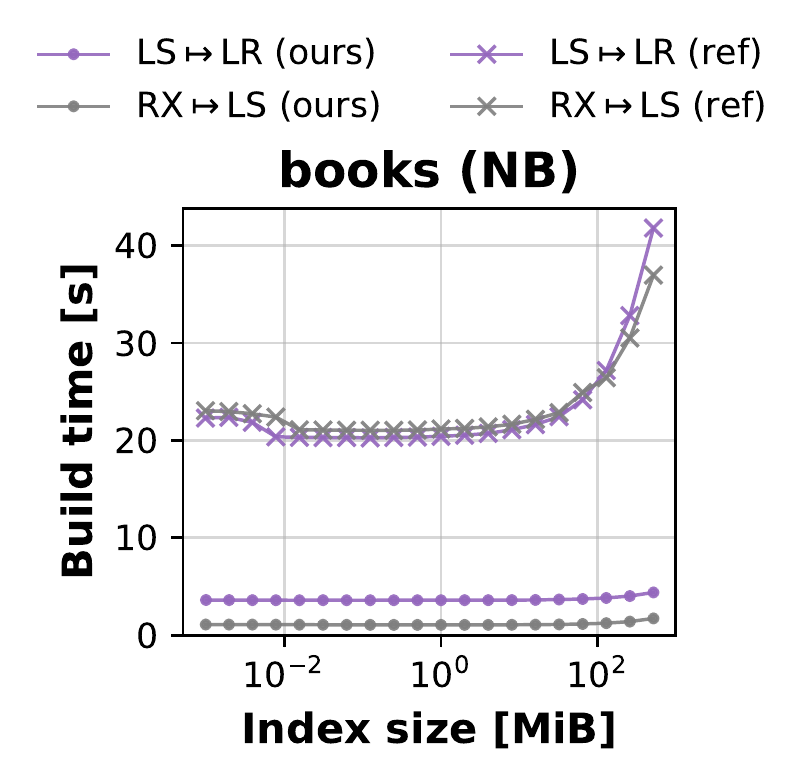}
        \caption{No bounds}
        \label{subfig:build:ref:none}
    \end{subfigure}
    \begin{subfigure}{.45\linewidth}
        \centering
        \includegraphics[scale=0.4]{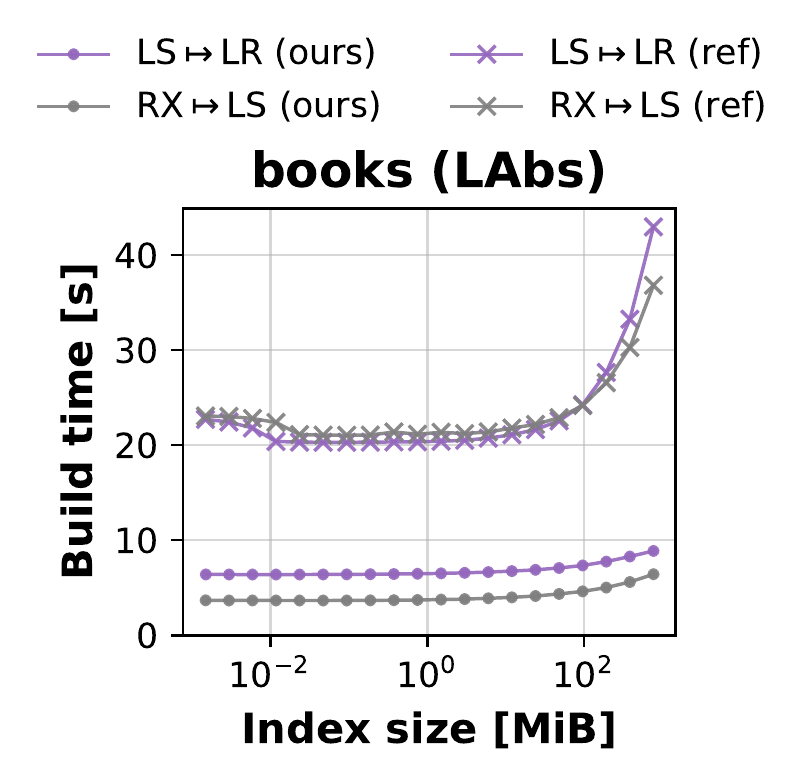}
        \caption{Local absolute bounds}
        \label{subfig:build:ref:labs}
    \end{subfigure}
    \caption{Build times of our implementation~(ours) and the reference implementation~(ref) with NB and LAbs.}
    \label{fig:build:ref}
\end{figure}

\noindent\textbf{Reference implementation.} \autoref{fig:build:ref} shows build times of our implementation (ours) and the reference implementation (ref). \autoref{subfig:build:ref:none} and \autoref{subfig:build:ref:labs} compare configurations with NB and LAbs, respectively. Build times for both types of bounds are almost identical for the reference implementation because the reference implementation always computes bounds during training and only decides later whether these computed bounds are kept or discarded. Considering only configurations with LAbs, our implementation improves build times by 2.5x to 6.3x. We attribute this improvement to our optimized segmentation for monotonous root models that avoids copying keys as described in \autoref{subsec:setup:impl}.

\noindent\textbf{Summary.}
RMIs can be built in a matter of seconds. For a given combination of models, the build time remains almost constant as long as the RMI fits in the cache. The computation of error bounds leads to additional cache and branch misses, which negatively impact build times.

\section{RMI Guideline}
\label{sec:guideline}

Based on our findings from the previous sections, we present a compact guideline for configuring RMIs. Our guideline does not guarantee to always provide the fastest lookups but it is easy to follow and achieves competitive lookup performance. Given a maximum allowed index size \emph{budget}, we propose to configure RMIs as follows.

\noindent\textbf{Models types.}
LS$\mapsto$LR with the maximum second-layer size that is allowed by the \emph{budget}. CS and LS both segment most datasets well, but we choose LS as it is slightly faster to train and evaluate. Although more accurate predictions can be obtained with CS, CS is only faster for small RMIs, where the improvement in search time outweighs the longer evaluation time. LR as second-layer model minimizes the error and thus always performs better than LS. Larger RMIs generally achieve smaller errors and thus perform better, which is why we choose the maximum number of second-layer models within the \emph{budget}.

\noindent\textbf{Error correction.}
LAbs+Bin or NB+MExp. Our experiments show that LAbs+Bin performs better than NB+MExp until a certain error threshold is reached. This error threshold is hardware-dependent and must be determined empirically once. We use the mean log$_2$ error as measure of error to estimate the number of search steps with exponential search and determine the error threshold to be~5.8. Whenever the mean log$_2$ error of our RMI with NB is below the threshold, we use NB+MExp and LAbs+Bin otherwise.

\begin{figure}
    \centering
    \includegraphics[scale=0.4]{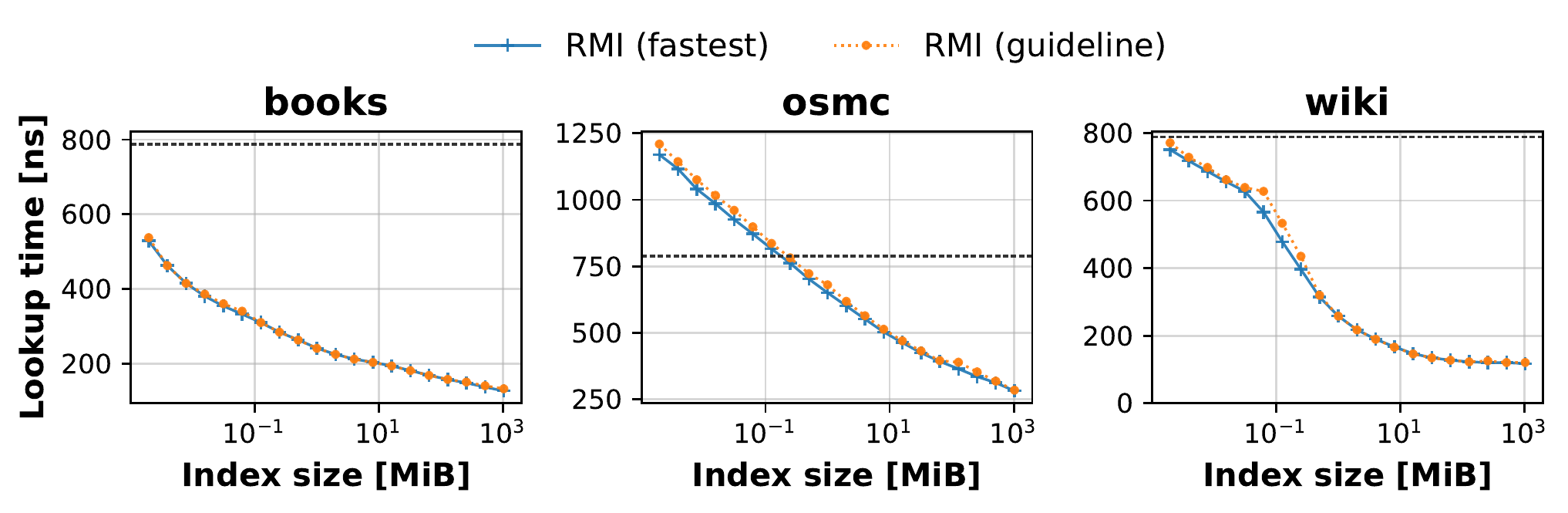}
    \caption{Comparison of lookup time of fastest configuration and guideline configuration.}
    \label{fig:guideline}
\end{figure}

\autoref{fig:guideline} compares the lookup times of configurations obtained by our guideline with the fastest configurations. As before, we omit \fb as none of the considered models segments \fb well. We consider size budgets between 2\,KiB and 1\,GiB. Our guideline is on average only 2.0\% slower than the fastest configuration with a maximum performance decline of 11.3\% on \wiki.

Implementing our guideline requires training at most two RMIs:
\begin{enumerate}[leftmargin=*]
    \item Train an RMI with LS$\mapsto$LR and NB that is within \emph{budget}.
    \item Compute the mean log$_2$ error of the RMI.
    \item If the error is above the threshold, train and use an RMI with LAbs within \emph{budget}. Otherwise, use the already trained RMI.
\end{enumerate}

\noindent\textbf{Limitations.}
In order to be simple and induce as little overhead as possible, our guideline neglects some aspects that are required for optimal configuration. (1)~Our guideline uses fixed model types. While LS$\mapsto$LR works well for datasets without outliers, a more suitable first-layer model must be sought for datasets with outliers. (2)~Our guideline only chooses between LAbs+Bin and NB+MExp based on a rough estimate of expected search steps. In some cases, other error correction strategies are slightly faster.

\section{Comparison with Other Indexes}
\label{sec:comparison}

In this section, we compare our guideline for configuring RMIs with the indexes introduced in \autoref{subsec:setup:baselines} and vary the parameters listed in \autoref{tab:setup:indexes} to obtain indexes of different sizes. Configurations of our RMI implementation are chosen based on our guideline. Configuration of the reference implementation are chosen based on its optimizer~\cite{marcus2020cdfshop}.

\subsection{Lookup Time}
\label{subsec:comparison:lookup}

\begin{figure}
    \centering
    \includegraphics[scale=0.4]{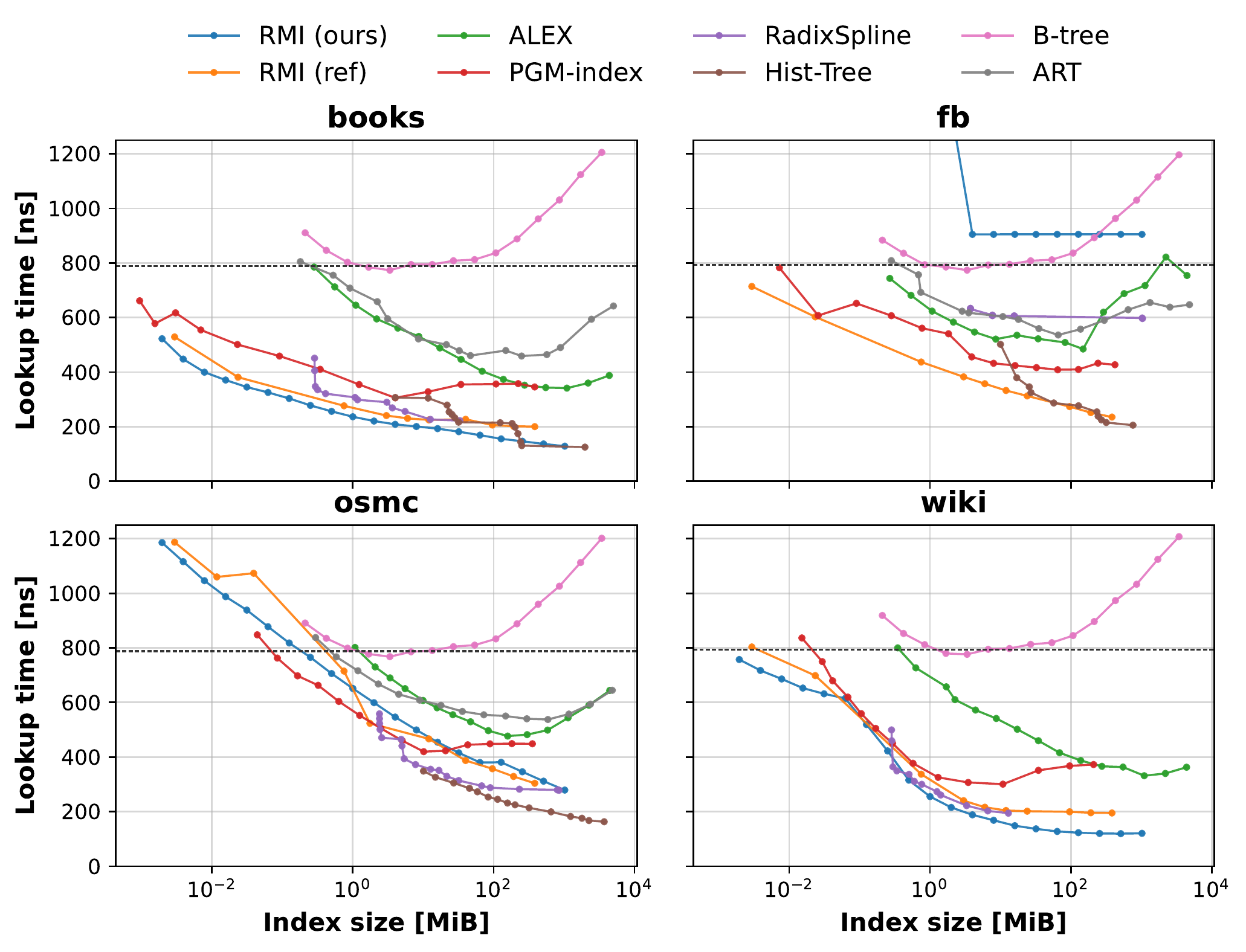}
    \caption{Comparison of lookup times w.r.t. index size.}
    \label{fig:comparison:lookup}
\end{figure}

We first compare lookup times with respect to index size. During a lookup, each index yields a search range, either through error bounds or level of sparsity. We use binary search to find keys in that search range. In \autoref{fig:comparison:lookup}, we report average lookup times. For indexes with multiple hyperparameters, i.e.,~RadixSpline and Hist-Tree, we show pareto-optimal configurations in terms of index size and lookup time for better readability. As a result, the number of data points shown differs across dataset. Hist-Tree and ART do not support duplicates and are therefore not evaluated on \wiki. Overall, our results are consistent with previous reports~\cite{kipf2019sosd,marcus2020benchmarking,learnedindexleaderboard}.

Let us first consider the traditional indexes. Hist-Tree is the fastest index on all datasets except \wiki, but Hist-Tree needs index sizes of 100\,MiB and more to reach its full potential. The best-performing configurations of Hist-Tree use a high branching factor resulting in few levels while achieving search intervals of less than 64 keys. B-tree is the only index whose performance is completely independent of the data distribution but also the slowest index, barely beating binary search. ART is always faster than B-tree but noticeably slower than all learned indexes except ALEX.

The performance of learned indexes highly depends on the data distribution. Up to a certain index size from which Hist-Tree outperforms the other indexes, learned indexes achieve the fastest lookup times. This implies that learned indexes work particularly well for smaller index sizes. On \books, \fb, and \wiki either our implementation of RMIs or the reference implementation dominates the other learned indexes. On \osmc, both PGM-index and RadixSpline perform better than RMIs. ALEX is clearly the slowest learned index, which can be attributed to its more complex adaptive structure.

Let us now compare our RMI implementation and the reference implementation~\cite{marcus2020cdfshop}. On \books and \wiki, our implementation dominates the reference implementation despite using our simple guideline. There are two reasons for this. (1)~Unlike how the optimizer is described~\cite{marcus2020cdfshop}, the publicly available implementation~\cite{rmicode} does not consider evaluation time in its optimization process and instead chooses configurations that achieve the smallest mean log$_2$ error. While this results in selecting the configuration with the fastest error correction time, it does not guarantee to select the configuration with the fastest lookup time. The configurations chosen by our guideline consistently have fast evaluation times at the cost of potentially slower error correction. (2)~The optimizer of the reference implementation always picks LAbs. Our experiments in \autoref{subsec:lookup:error_correction} show that for accurate RMIs, NB+MExp performs better, which is considered by our guideline. On \osmc, no implementation dominates the other. Here, RMIs are never sufficiently accurate for our guidelines to deviate from LAbs+Bin. Thus, differences in performance are solely due to the choice of models. On \fb, the reference implementation clearly dominates our implementation. As discussed before, LS is not sufficient hor segmenting datasets with extreme outliers. Here, the reference implementation chooses a variant of LR that ignores the lowest and highest 0.01\% of keys for segmentation. This approach, while effectively eliminating the outliers in \fb from the segmentation process, only works if there are at most 0.01\% of outliers at either end of the key space. We did not include this model type in our evaluation because we believe that a more robust solution to segmentation should be sought.

\begin{figure}
    \centering
    \includegraphics[scale=0.4]{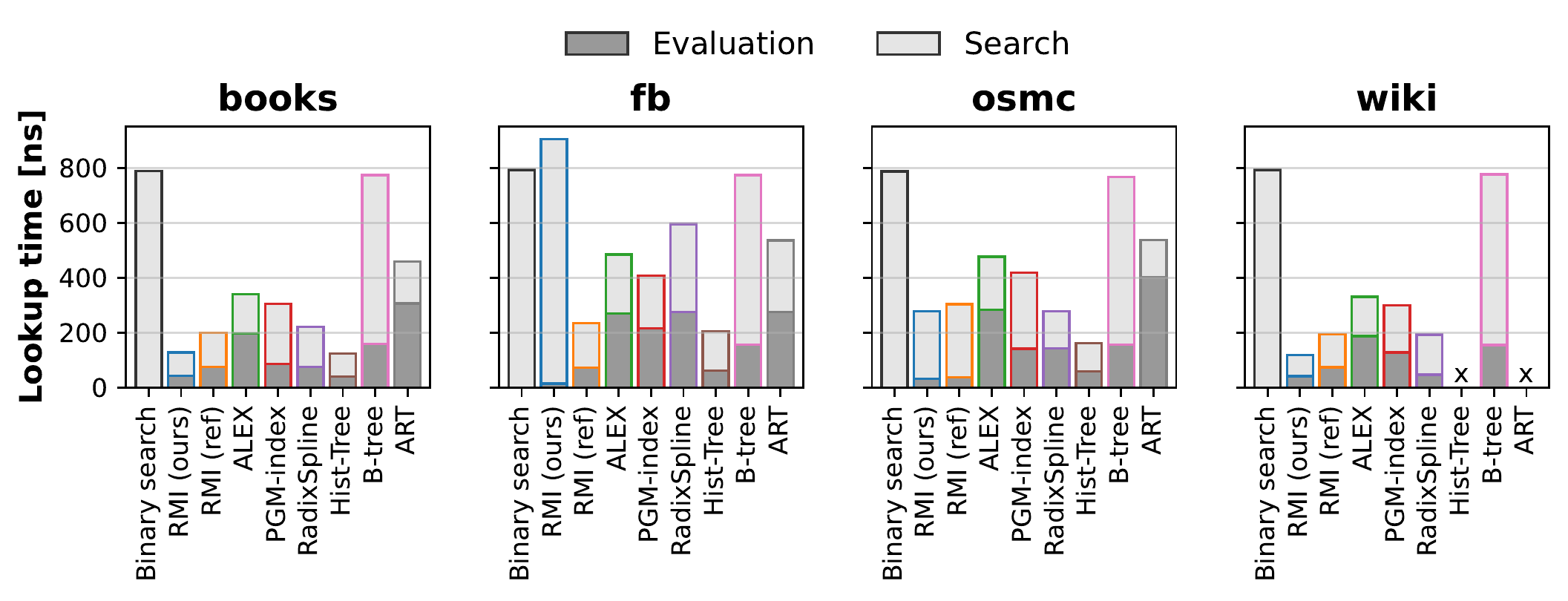}
    \caption{Comparison of evaluation time and search time for the best-performing configuration of each index.}
    \label{fig:comparison:eval_search}
\end{figure}

Let us now examine the composition of lookup time from evaluation time (evaluating the model or traversing the tree) and search time (searching within the error interval or data page). \autoref{fig:comparison:eval_search} shows the lookup time of the best-performing configuration of each index divided into evaluation time and search time. There is a trade-off between fast evaluation and fast search. RMIs clearly prioritize fast evaluation: The evaluation leads to the correct segment in a fixed number of steps, but the RMI does not provide any guarantees on the prediction accuracy. Adding more segments continuously improves the lookup performance because more segments hardly increase the evaluation time while improving the search time. If the evaluation time of our implementation is faster than that of the reference implementation, it is because our configuration does not use bounds. In contrast, PGM-index and RadixSpline prioritize fast error correction: Both indexes cap the maximum error at the cost of a slower evaluation that requires traversing multiple layers or performing intermediate searches. At some point, the improved search time of a smaller maximum error does not compensate the longer evaluation time and the lookup performance decreases. Thus, despite fewer hyperparameters than RMIs, configuring PGM-index and RadixSpline optimally is an elaborate task.

\noindent\textbf{Summary.}
Learned indexes perform well even at small index sizes. Overall, Hist-Tree is the fastest evaluated index, but it requires sizes of 100\,MiB and more to beat learned indexes. Other traditional indexes perform significantly worse on sorted data.

\subsection{Build Time}
\label{subsec:comparison:build}

\begin{figure}
    \centering
    \includegraphics[scale=0.4]{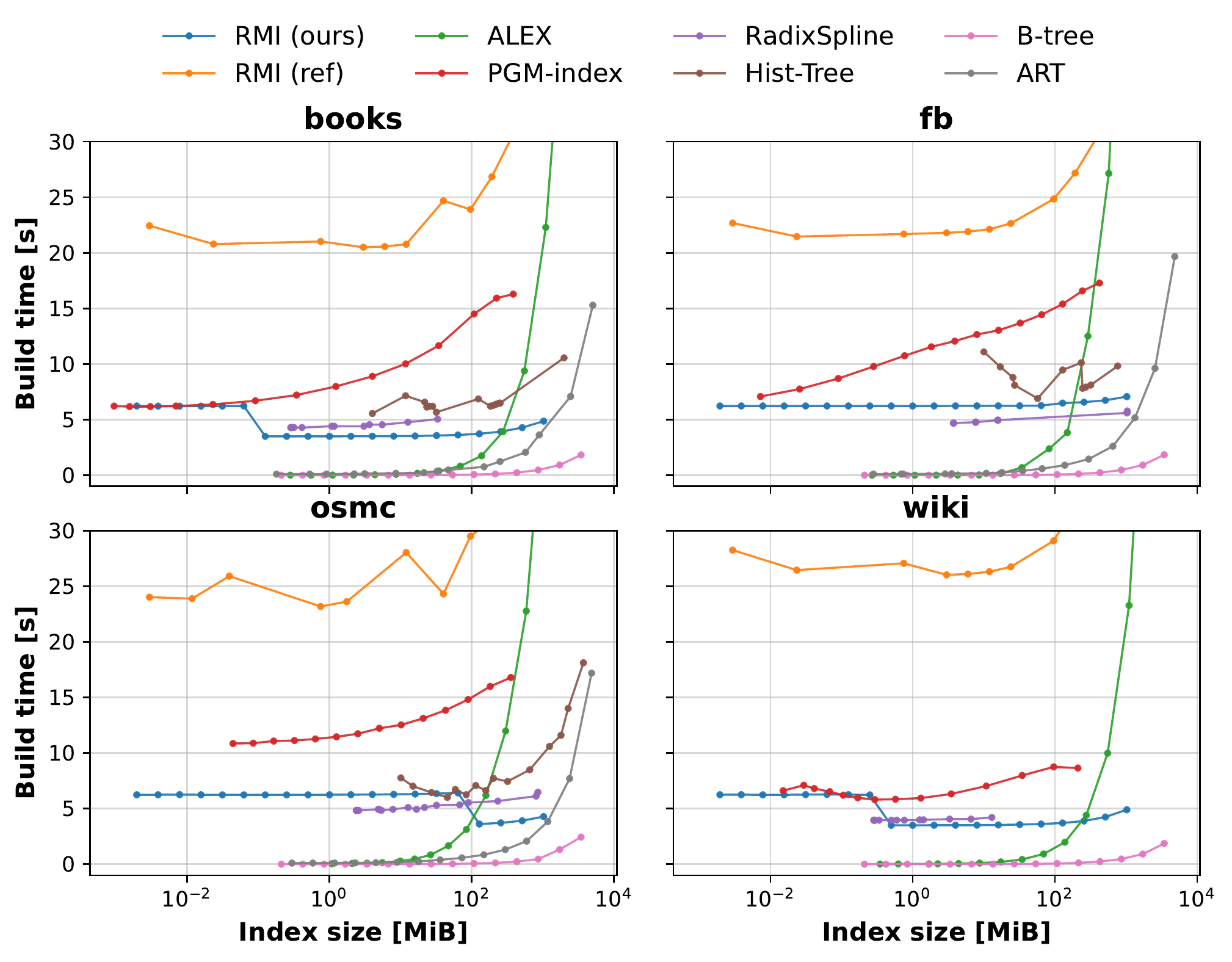}
    \caption{Comparison of build times w.r.t. index size.}
    \label{fig:comparison:build}
\end{figure}

Next, we compare build times with respect to index size. In \autoref{fig:comparison:build}, we report build times which refer to the index configurations evaluated in terms of lookup time in \autoref{subsec:comparison:lookup}. We show the raw build times without the time required to determine hyperparameters, e.g.,~by running the reference implementation's optimizer~\cite{marcus2020cdfshop} or determining pareto-optimal configurations of RadixSpline and Hist-Tree. Some indexes require data preparation to be built. For instance, ALEX, B-tree, and ART are not only built on the keys but also explicitly require the positions to which these keys should be mapped. Since these preparation steps could be circumvented by a specialized implementation, we do not consider it part of the build time.

The index size of B-tree, ART, and ALEX is determined by the level of sparsity. In contrast to learned indexes, these indexes are built on a subset of the keys and therefore provide fast build times especially at smaller index sizes. With an increasing number of keys, the structure of these indexes becomes more complex, e.g.,~more levels are introduced, and the build time increases. In contrast, RMI, PGM-index, and RadixSpline are always built on the entire dataset. This means that their build times are higher from the outset. RMI and RadixSpline have a fixed number of layers. Therefore, their build time is hardly impacted by the data distribution and only increases once the index no longer fits into cache. The sudden decrease in build time of RMIs on \books and \wiki is caused by the guideline choosing an RMI configuration without bounds which is faster to build. PGM-index, on the other hand, has a variable number of layers. Depending on the data distribution and the desired error, more layers have to be trained leading to a steeper increase in build times compared to RadixSpline and RMI. Causes for the differences in build time between our RMI implementation and the reference implementation~\cite{marcus2020cdfshop} were already discussed in \autoref{sec:build}. The reference implementation's jumps in build time are caused by varying build times for different model types chosen by its optimizer. Hist-Tree exhibits similar build times to the learned indexes. However at larger sizes, its built time quickly increases due to the increasing depth of the Hist-Tree.

\noindent\textbf{Summary.}
The benefits in terms of lookup performance of learned indexes come at the cost of significantly higher build times compared to traditional indexes. Thus, the improvement of build times should be a priority of future work.

\section{Conclusion and Future Work}
\label{sec:conclusion}

We provided an extensible open-source implementation of RMIs and conducted a comprehensive hyperparameter analysis of RMIs in terms of prediction accuracy, lookup time, and build time. Based on this analysis, we developed a simple-to-follow guideline for configuring RMIs, which achieves competitive performance. In addition, we were able to improve the build time of RMIs by exploiting the monotonicity of models, thereby avoiding the copying of keys when assigning them to the second-layer models. In the future, we plan to extend our implementation to also support multi-layer RMIs and additional model types. We would also like to address the problem of segmenting datasets with extreme outliers.

\balance

\bibliographystyle{ACM-Reference-Format}
\bibliography{main}

\appendix

\section{Complete Experimental Results}
\label{sec:results}

\begin{figure*}
    \centering
    \includegraphics[scale=0.37]{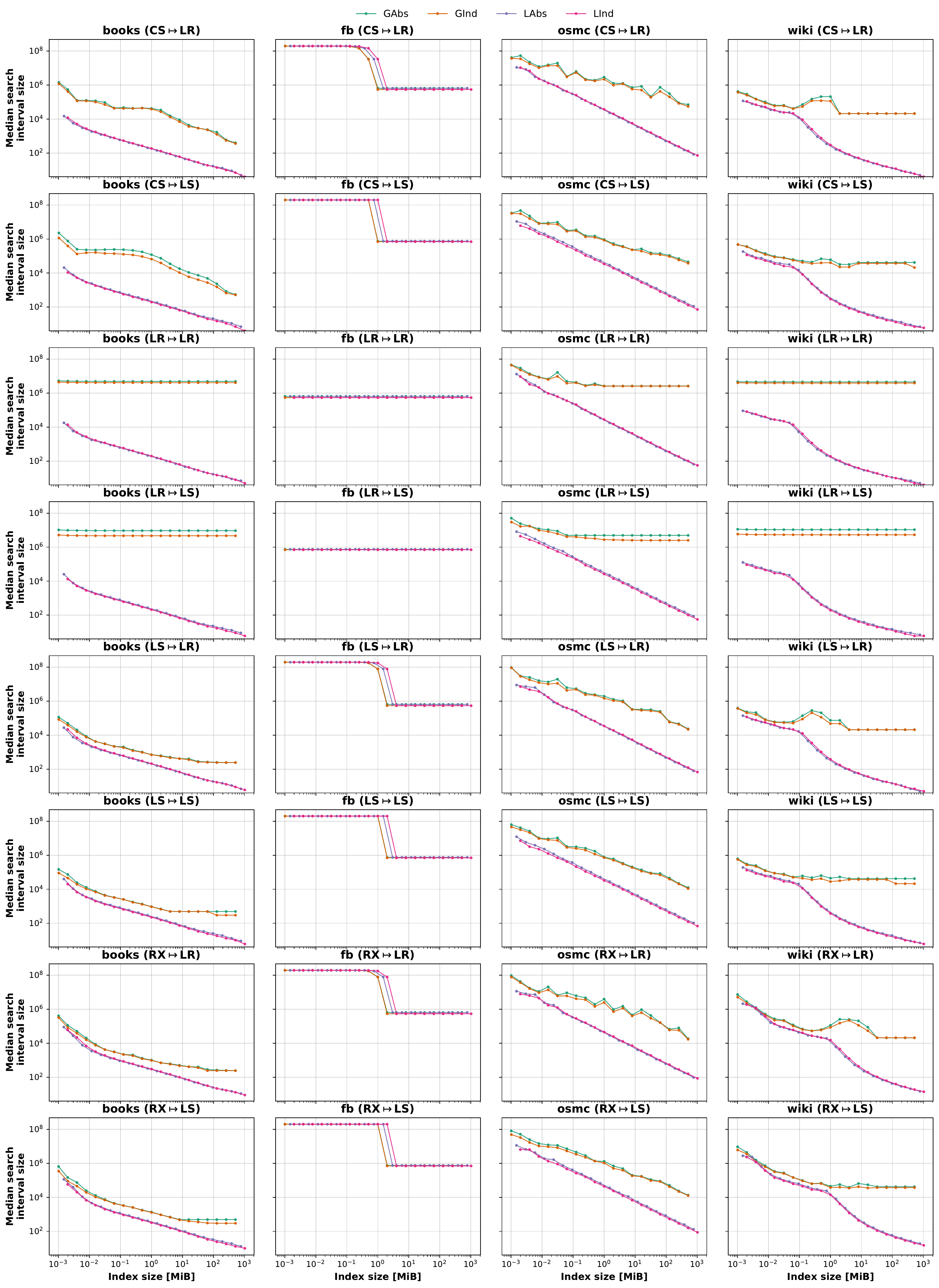}
    \caption{Comparison of error interval sizes for different error bounds.}
    \label{fig:appendix:error:median-interval}
\end{figure*}

\begin{figure*}
    \centering
    \includegraphics[scale=0.37]{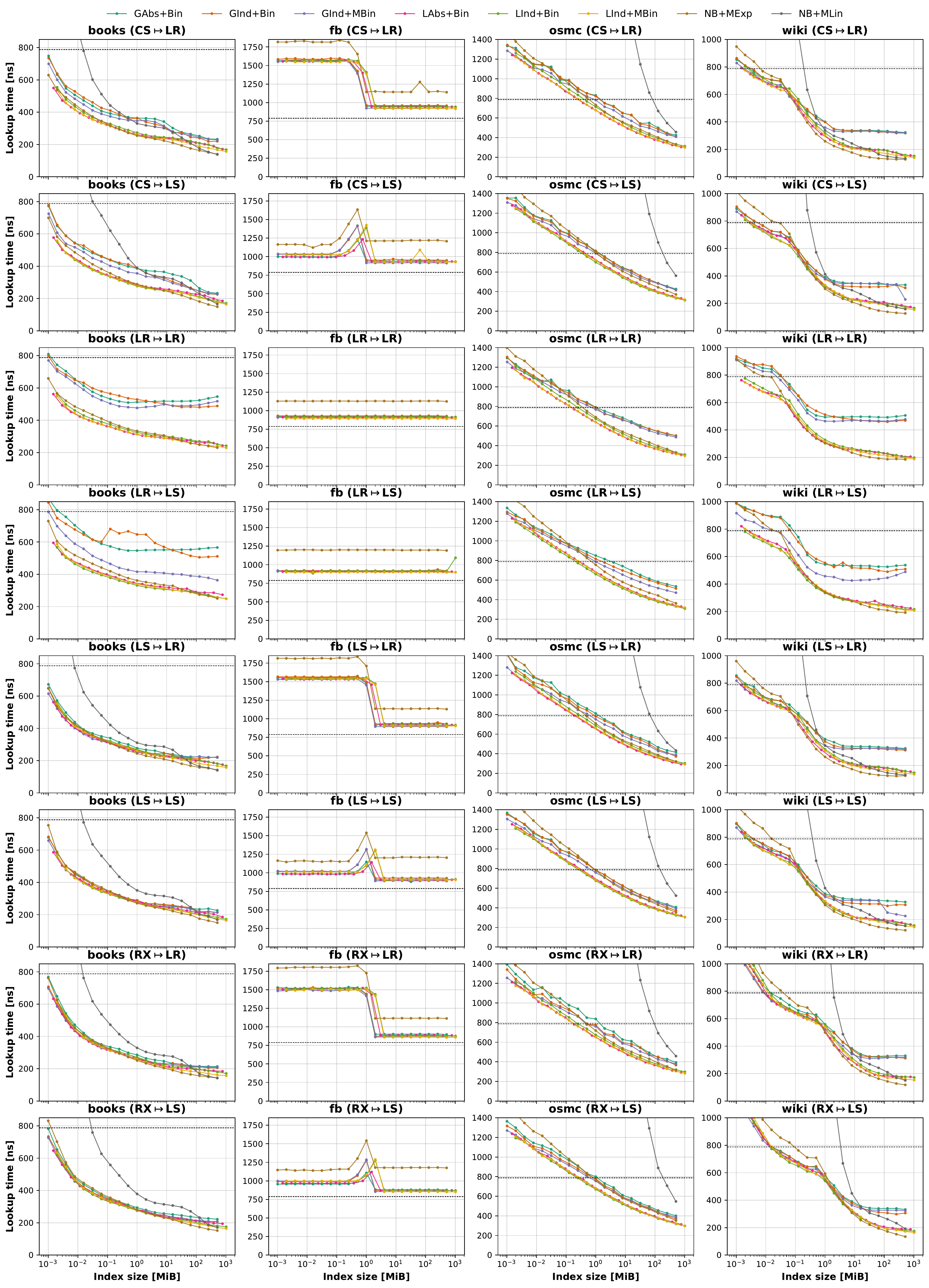}
    \caption{Comparison of lookup time for different error correction strategies.}
    \label{fig:appendix:lookup:error-correction}
\end{figure*}

\begin{figure*}
    \centering
    \includegraphics[scale=0.37]{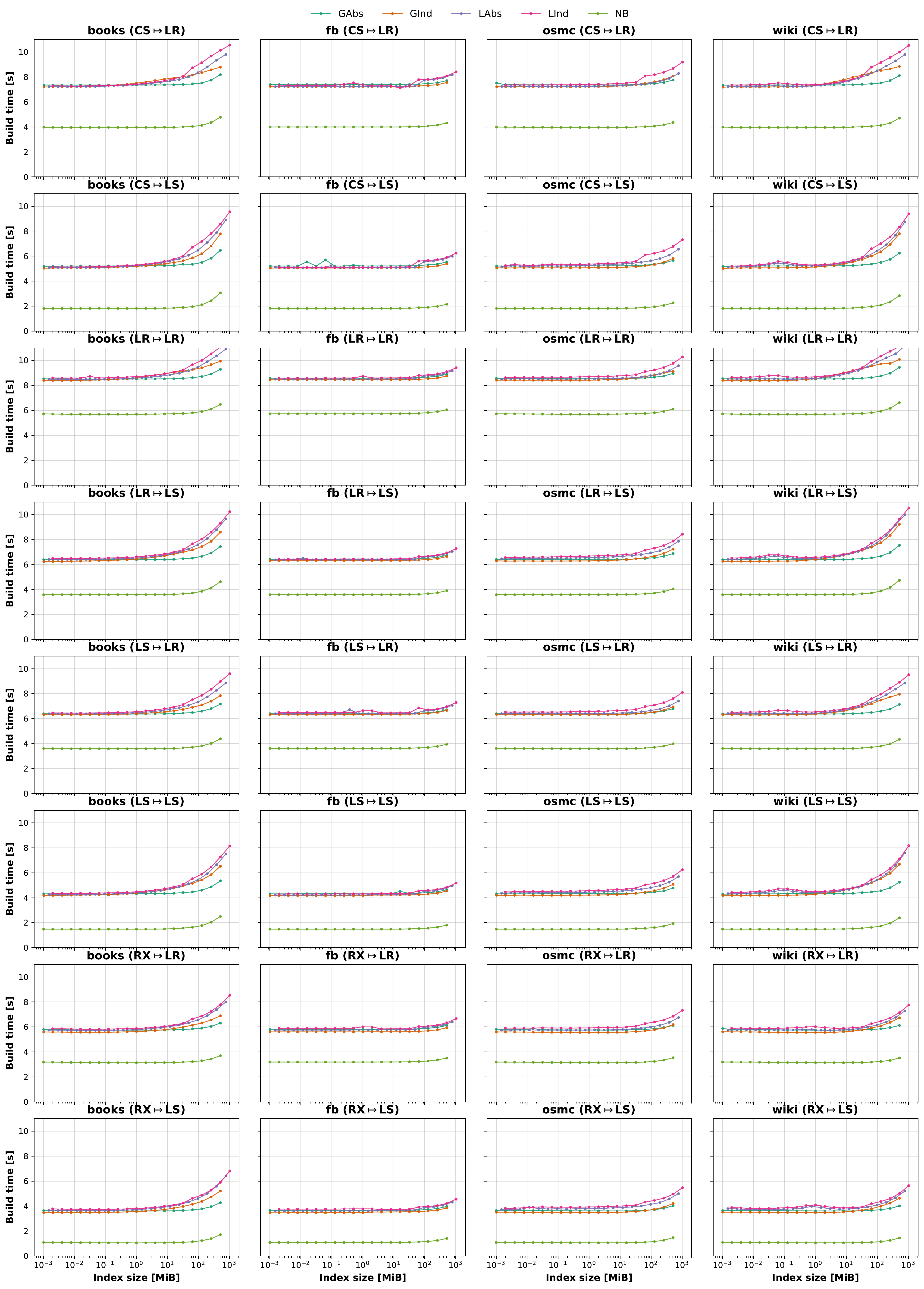}
    \caption{Comparison of build time for different error bounds.}
    \label{fig:appendix:build:ours:bounds}
\end{figure*}

Complete experimental results on the impact of error bounds on error interval sizes as introduced in \autoref{subsec:error:bounds} are shown in \autoref{fig:appendix:error:median-interval}.
Complete experimental results on the impact of error bounds and search algorithm on lookup time as introduced in \autoref{subsec:lookup:error_correction} are shown in \autoref{fig:appendix:lookup:error-correction}.
Complete experimental results on the impact of model type and error bounds on build time as introduced in \autoref{sec:build} are shown in \autoref{fig:appendix:build:ours:bounds}.

\end{document}